\documentclass[12pt,a4paper]{article}%
\usepackage{amsfonts}
\usepackage{amsmath}
\usepackage{amssymb}
\usepackage{lineno}
\usepackage{float}
\usepackage{algorithm}
\usepackage{hyperref}
\usepackage[numbers,square,sectionbib]{natbib}%
\usepackage{graphicx}
\providecommand{\U}[1]{\protect\rule{.1in}{.1in}}
\ifx\pdfoutput\relax\let\pdfoutput=\undefined\fi
\newcount\msipdfoutput
\ifx\pdfoutput\undefined\else
\ifcase\pdfoutput\else
\ifx\paperwidth\undefined\else
\ifdim\paperheight=0pt\relax\else\pdfpageheight\paperheight\fi
\ifdim\paperwidth=0pt\relax\else\pdfpagewidth\paperwidth\fi
\fi\fi\fi
\begin{document}

\begin{center}
\textbf{The Weibull Birnbaum-Saunders Distribution: Properties and
Applications}\medskip\medskip\medskip\medskip

\textbf{Lazhar Benkhelifa}\medskip\medskip\medskip\medskip

\textit{Laboratory of Applied Mathematics, Mohamed Khider University, Biskra,
07000, Algeria\medskip\medskip}

\textit{Departement of Mathematics and \textit{I}nformatics, Larbi Ben M'Hidi
University, Oum El Bouaghi, 04000, Algeria\bigskip}

l.benkhelifa@yahoo.fr\textit{\bigskip}
\end{center}

\noindent\textbf{Abstract}\medskip

\noindent This paper introduces a new four-parameter lifetime model called the
Weibull Birnbaum-Saunders distribution. This new distribution represents a
more flexible model for the lifetime data. Its failure rate function can be
increasing, decreasing, upside-down bathtub shaped, bathtub-shaped or modified
bathtub shaped depending on its parameters. Some structural properties of the
proposed model are investigated including expansions for the cumulative and
density functions, moments, generating function, mean deviations, order
statistics and reliability. The maximum likelihood estimation method is used
to estimate the model parameters and the observed information matrix is
determined. The flexibility of the new model is shown by means of two real
data sets\textbf{.\bigskip}\bigskip


\bigskip

\section{\textbf{Introduction\label{sec 1}}}

The two-parameter Birnbaum--Saunders (BS) distribution, which was introduced
by Birnbaum and Saunders \cite{r3}, is a very important lifetime model and is
widely used in reliability studies. This distribution, also known as the
fatigue life distribution, was originally derived from a model that shows the
total time that passes until that some type of cumulative damage, produced by
the development and growth of a dominant crack, surpasses a threshold value
and causes the material specimen to fail. Desmond \cite{r10} provided a more
general derivation based on a biological model and strengthened the physical
justification for the use of this distribution.\medskip

\noindent A random variable $T$ having\textbf{\ }the BS distribution\ with
shape parameter $\alpha>0$ and scale parameter $\beta>0$, denoted by $T\sim$
BS$\left(  \alpha,\beta\right)  ,$ is defined by%
\[
T=\beta\left[  \frac{\alpha Z}{2}+\sqrt{\left(  \frac{\alpha Z}{2}\right)
^{2}+1}\right]  ^{2},
\]
where $Z$ is a standard normal random variable. The cumulative distribution
function (cdf) of $T$ is%
\begin{equation}
G\left(  t\right)  =\Phi\left(  v\right)  ,\text{ }t>0, \label{1}%
\end{equation}
where $\Phi\left(  \cdot\right)  $ is the standard normal distribution
function, $v=\alpha^{-1}\rho\left(  t/\beta\right)  $ and $\rho\left(
z\right)  =z^{1/2}-z^{-1/2}.$ The probability density function (pdf)
corresponding to (\ref{1}) is given by%
\begin{equation}
g\left(  t\right)  =\kappa\left(  \alpha,\beta\right)  t^{-3/2}\left(
t+\beta\right)  \exp\left\{  -\frac{\tau\left(  t/\beta\right)  }{2\alpha^{2}%
}\right\}  ,\text{ }t>0, \label{2}%
\end{equation}
where $\kappa\left(  \alpha,\beta\right)  =\exp\left(  \alpha^{-2}\right)
/\left(  2\alpha\sqrt{2\pi\beta}\right)  $ and $\tau\left(  z\right)
=z+z^{-1}.$ The fractional moments of $T$ are given by (see \cite{r20})%
\[
E\left(  T^{k}\right)  =\beta^{k}I\left(  \alpha,\beta\right)  ,
\]
where%
\begin{equation}
I\left(  \alpha,\beta\right)  =\frac{K_{k+1/2}\left(  \alpha^{-2}\right)
+K_{k-1/2}\left(  \alpha^{-2}\right)  }{2K_{1/2}\left(  \alpha^{-2}\right)  },
\label{33}%
\end{equation}
and the function $K_{\nu}\left(  z\right)  $ denotes the modified Bessel
function of the third kind with $\nu$ representing its order and $z$ the
argument. The parameter $\beta$ is the median of the BS distribution, because
$G\left(  \beta\right)  =\Phi\left(  0\right)  =1/2.$\medskip

\noindent Since the BS distribution was proposed, it has received much
attention in the literature. This attention for the BS distribution is due to
its many attractive properties and its relation to the normal distribution.
For more details on the BS distribution, we refer to \cite{r15} and the
references therein. The BS distribution has been used in several research
areas such as engineering, environmental sciences, finance, and wind energy.
However, it allows for upside-down hazard rates only (see \cite{r14}), hence
cannot provide reasonable fits for modeling phenomenon with decreasing,
increasing, modified bathtub shaped and bathtub-shaped failure rates which are
common in reliability studies.\medskip

\noindent For this reason, several generalizations and extensions of the BS
distribution have been proposed in the literature. For example, Cordeiro and
Lemonte \cite{r6}, using the beta-G class \cite{r11}, proposed an extension of
BS distribution named as the beta BS distribution.\ Saulo et al. \cite{r22},
based on the work of Cordeiro and de Castro \cite{r5}, defined the Kumaraswamy
BS distribution. Lemonte \cite{r16}, based on the scheme introduced by
Marshall and Olkin \cite{r17}, deifined the Marshall--Olkin extended BS
distribution. Cordeiro et al. \cite{r7} adopted the McDonald-G class \cite{r2}
to define the McDonald BS distribution. Cordeiro et al. \cite{r8} used the
generator approach of Zografos and Balakrishnan \cite{r28} to introduce the
gamma Birnbaum-Saunders distribution. In this paper, a new four-parameter
extension for the BS distribution is proposed.\medskip

\noindent Recently, Bourguignon et al. \cite{r4} proposed an interesting
method of adding a new parameter to an existing $G$ distribution. The
resulting distribution, known as the Weibull-G distribution, gives more
flexibility to model various types of data. Let $G\left(  t,\mathbf{\theta
}\right)  \ $be a continuous baseline distribution with density $g$ depends on
a parameter vector $\mathbf{\theta}$ and the Weibull cdf $F_{W}\left(
w\right)  =1-e^{-aw^{b}}$ (for $w>0$) with positive parameters $a$ and $b$.
Bourguignon et al. \cite{r4} replaced the argument $w$ by $G\left(
w,\mathbf{\xi}\right)  /\overline{G}\left(  w,\mathbf{\theta}\right)  $ where
$\overline{G}\left(  w,\mathbf{\xi}\right)  =1-G\left(  w,\mathbf{\theta
}\right)  $, and defined the cdf of their class by$\ $%
\begin{equation}
F\left(  t;a,b,\mathbf{\theta}\right)  =ab\int_{0}^{\left[  \frac{G\left(
t,\mathbf{\theta}\right)  }{\overline{G}\left(  t,\mathbf{\theta}\right)
}\right]  }w^{b-1}e^{-aw^{b}}dw=1-\exp\left(  -a\left[  \frac{G\left(
t,\mathbf{\theta}\right)  }{\overline{G}\left(  t,\mathbf{\theta}\right)
}\right]  ^{b}\right)  . \label{3}%
\end{equation}
Then, the Weibull-G density function is given by%
\begin{equation}
f\left(  t;a,b,\mathbf{\theta}\right)  =abg\left(  t,\mathbf{\theta}\right)
\left[  \frac{G\left(  t,\mathbf{\theta}\right)  ^{b-1}}{\overline{G}\left(
t,\mathbf{\theta}\right)  ^{b+1}}\right]  \exp\left(  -a\left[  \frac{G\left(
t,\mathbf{\theta}\right)  }{\overline{G}\left(  t,\mathbf{\theta}\right)
}\right]  ^{b}\right)  . \label{4}%
\end{equation}
Some generalized distributions have been proposed under this methodology.
Tahir et al. \cite{r23, r24, r25} defined the Weibull-Pareto, Weibull-Lomax
and Weibull-Dagum distributions by taking $G\left(  t,\mathbf{\theta}\right)
$ to be the cdf of the Pareto, Lomax and Dagum distributions, respectively.
More recently, Afify et al. \cite{r1} defined and studied the Weibull
Fr\'{e}chet distribution. In a similar way, we propose a new extension for the
BS distribution called the\ Weibull BS (WBS) distribution, which has been
applied to the modeling of fatigue failure times and reliability
studies.\medskip

\noindent The rest of the paper is organized as follows. In Section
\ref{sec 2}, we introduce the WBS distribution and plot its density and
failure rate functions. In Section \ref{sec 3}, we provide a mixture
representation for its density and cumulative distributions. Structural
properties such as the ordinary moments, generating function, quantile
function and simulation, mean deviations, the density of the order statistics
and the reliability are derived in Section \ref{sec4}. In Section \ref{sec5},
we discuss maximum likelihood estimation of the WBS parameters and derive the
observed information matrix. Two applications are presented in Section
\ref{sec6} to show the potentiality of the new distribution. Some concluding
remarks are given in Section \ref{sec7}.

\section{\textbf{The WBS distribution\label{sec 2}}}

Substituting (\ref{1}) in (\ref{3}), the cdf of the WBS distribution can be
written as%
\begin{equation}
F\left(  t\right)  =1-\exp\left(  -a\left[  \frac{\Phi\left(  v\right)
}{1-\Phi\left(  v\right)  }\right]  ^{b}\right)  . \label{5}%
\end{equation}
The pdf corresponding to (\ref{5}) is given by%
\begin{align}
f\left(  t\right)   &  =ab\kappa\left(  \alpha,\beta\right)  t^{-3/2}\left(
t+\beta\right)  \exp\left(  -\frac{\tau\left(  t/\beta\right)  }{2\alpha^{2}%
}\right) \nonumber\\
&  \times\left[  \frac{\Phi\left(  v\right)  ^{b-1}}{\left\{  1-\Phi\left(
v\right)  \right\}  ^{b+1}}\right]  \exp\left(  -a\left[  \frac{\Phi\left(
v\right)  }{1-\Phi\left(  v\right)  }\right]  ^{b}\right)  , \label{6}%
\end{align}
where $\beta$ is a scale parameter and $\alpha$, $a$ and $b$ are positive
shape parameters. It is clear that the BS distribution is not a special case
of WBS distribution. If a random variable $T$ follows a WBS distribution with
parameters $\alpha,\beta,a\ $and $b$ will be denoted by $T\sim$ WBS$\left(
\alpha,\beta,a,b\right)  .$ The reliability and the failure rate function of
$T$ are, respectively, given by%
\[
R(t)=\exp\left(  -a\left[  \frac{\Phi\left(  v\right)  }{1-\Phi\left(
v\right)  }\right]  ^{b}\right)  ,
\]
and%
\[
h(t)=ab\kappa\left(  \alpha,\beta\right)  t^{-3/2}\left(  t+\beta\right)
\left[  \frac{\Phi^{b-1}\left(  v\right)  }{\left\{  1-\Phi\left(  v\right)
\right\}  ^{b+1}}\right]  \exp\left\{  -\frac{\tau\left(  t/\beta\right)
}{2\alpha^{2}}\right\}  .
\]
Plots of pdf and failure rate function of the WBS distribution for selected
values of the parameters are given in Figure 1 and Figure 2, respectively.
Figure 1 indicates that the WBS pdf can take various shapes such as symmetric,
right-skewed and left-skewed depending on the parameter values. Figure 2 shows
that the failure rate function of the WBS distribution can be increasing,
decreasing, upside-down bathtub (unimodal) shaped, bathtub-shaped or modified
bathtub shaped (unimodal shape followed by increasing) depending on the
parameter values. So, the WBS distribution is quite flexible and can be used
effectively in analyzing survival data.%

%TCIMACRO{\FRAME{itbpFU}{5.1508in}{2.7752in}{0in}{\Qcb{Figure 1. Plots of the
%WBS pdf for some values of the parameters.}}{}{fig1.eps}%
%{\special{ language "Scientific Word";  type "GRAPHIC";  display "USEDEF";
%valid_file "F";  width 5.1508in;  height 2.7752in;  depth 0in;
%original-width 18.9749in;  original-height 9.1584in;  cropleft "0";
%filename 'fig1.eps';file-properties "XNPEU";}} }%
%BeginExpansion
{\parbox[b]{5.1508in}{\begin{center}
\ifcase\msipdfoutput
\includegraphics[
height=2.7752in,
width=5.1508in
]%
{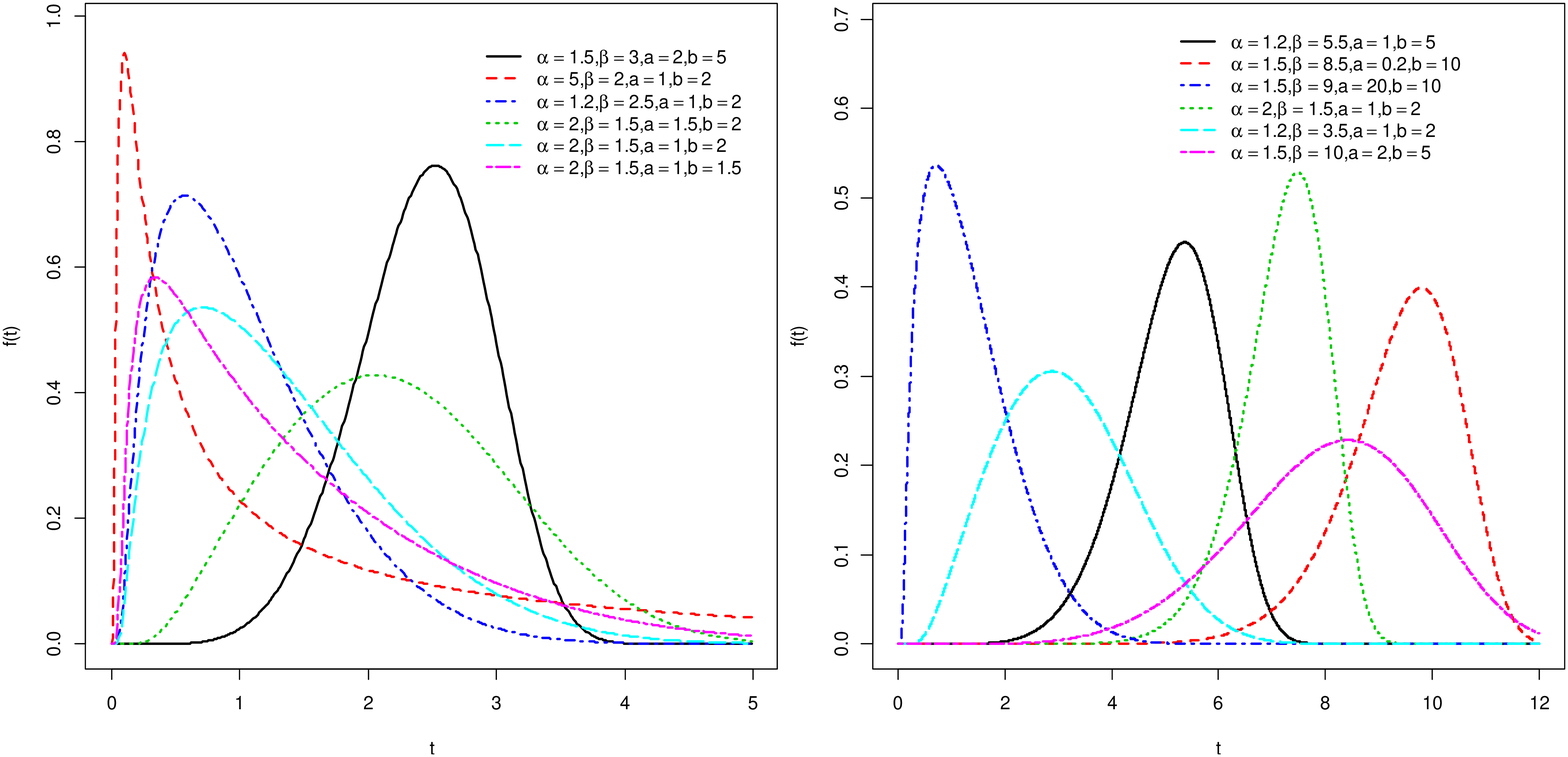}%
\else
\includegraphics[
height=2.7752in,
width=5.1508in
]%
{C:/Users/LAZHAR/IDEES/Weibull-Birnbaum-Saunders distribution/arXiv/graphics/fig1.jpg}%
\fi
\\
Figure 1. Plots of the WBS pdf for some values of the parameters.
\end{center}}}
%EndExpansion
%

%TCIMACRO{\FRAME{itbpFU}{5.1439in}{2.7752in}{0in}{\Qcb{Figure 2. Plots of the
%WBS failure rate function for some values of the parameters.}}{}%
%{fig2.eps}{\special{ language "Scientific Word";  type "GRAPHIC";
%display "USEDEF";  valid_file "F";  width 5.1439in;  height 2.7752in;
%depth 0in;  original-width 18.9749in;  original-height 9.1584in;
%cropleft "0";  croptop "1";  cropright "1";  cropbottom "0";
%filename 'fig2.eps';file-properties "XNPEU";}} }%
%BeginExpansion
{\parbox[b]{5.1439in}{\begin{center}
\ifcase\msipdfoutput
\includegraphics[
height=2.7752in,
width=5.1439in
]%
{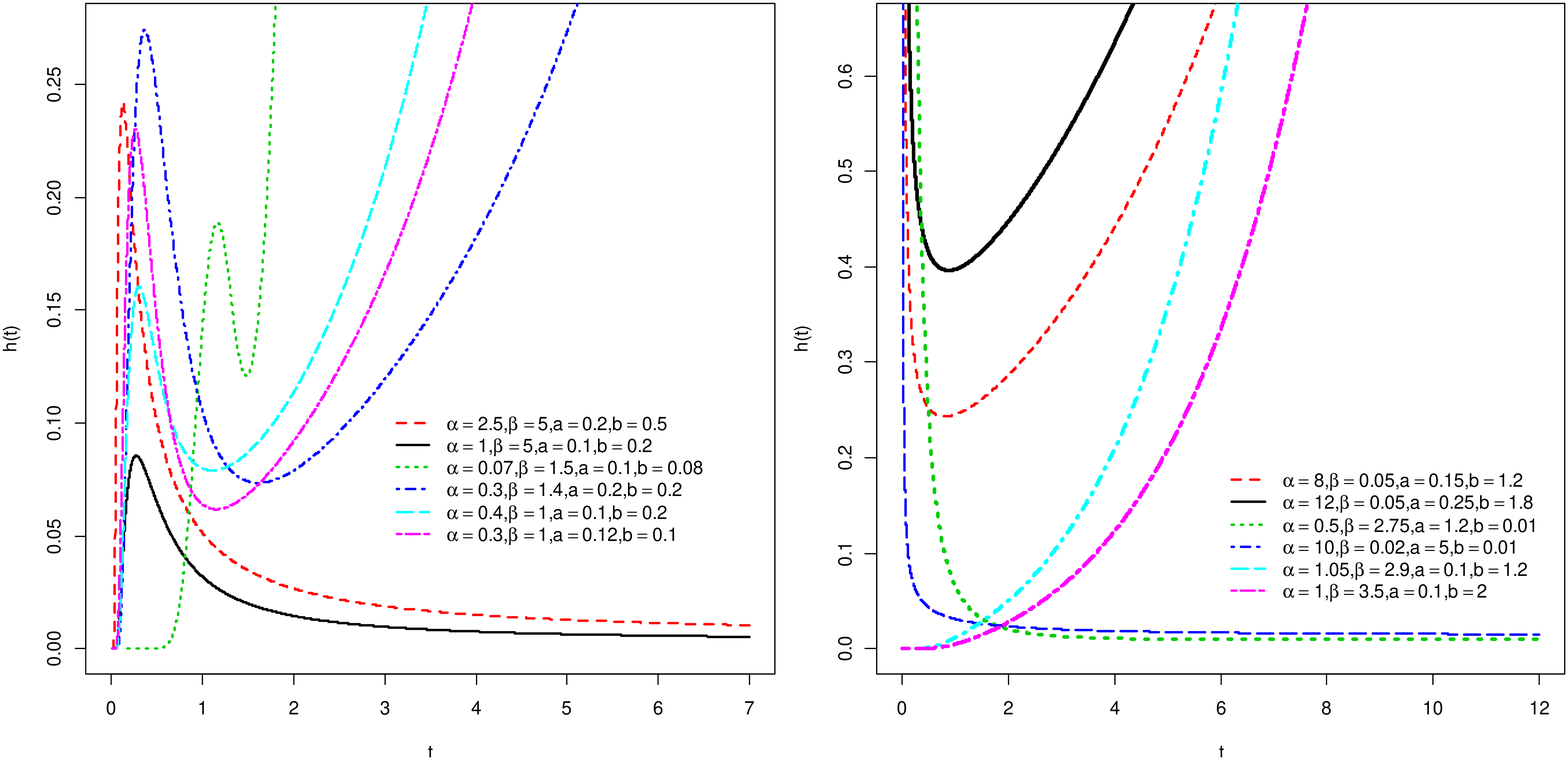}%
\else
\includegraphics[
height=2.7752in,
width=5.1439in
]%
{C:/Users/LAZHAR/IDEES/Weibull-Birnbaum-Saunders distribution/arXiv/graphics/fig2.jpg}%
\fi
\\
Figure 2. Plots of the WBS failure rate function for some values of the
parameters.
\end{center}}}
%EndExpansion

\section{\textbf{Mixture representation\label{sec 3}}}

In this section, we derive expansions for the pdf and cdf of the WBS
distribution. The pdf\ and cdf of the WBS distribution can be written as a
linear combination of the pdf and cdf of exponentiated BS (EBS) distribution,
respectively. A random variable $X$ having the EBS distribution with
parameters $\alpha$, $\beta$ and $a>0$, denoted by $X\sim$ EBS$\left(
\alpha,\beta,a\right)  $, if its cdf and pdf are given by $H\left(  x\right)
=\Phi^{a}\left(  v\right)  $ and $h(x)=ag(x)\Phi^{a-1}(v)$, respectively,
where $v$ is defined in (\ref{1}) and $g$ is given in (\ref{2}). The
properties of exponentiated distributions have been studied by several
authors. For example, Mudholkar and Srivastava \cite{r19} studied the
exponentiated Weibull distribution, Gupta and Kundu \cite{r13} studied the
exponentiated exponential distribution and Sarhan and Apaloo \cite{r21}
proposed the exponentiated modified Weibull extension distribution.\medskip

\noindent The pdf of the WBS distribution (\ref{6}) can be written as%
\begin{equation}
f\left(  t\right)  =abg\left(  t\right)  \frac{\Phi^{b-1}\left(  v\right)
}{\left[  1-\Phi\left(  v\right)  \right]  ^{b+1}}\exp\left(  -a\left[
\frac{\Phi\left(  v\right)  }{1-\Phi\left(  v\right)  }\right]  ^{b}\right)  .
\label{7}%
\end{equation}
Using the series expansion for the exponential function, we obtain%
\begin{equation}
\exp\left(  -a\left[  \frac{\Phi\left(  v\right)  }{1-\Phi\left(  v\right)
}\right]  ^{b}\right)  =\sum_{k=0}^{\infty}\frac{\left(  -1\right)  ^{k}a^{k}%
}{k!}\frac{\Phi^{bk}\left(  v\right)  }{\left[  1-\Phi\left(  v\right)
\right]  ^{bk}}. \label{8}%
\end{equation}
Substituting (\ref{8}) in (\ref{7}), we get%
\[
f\left(  t\right)  =abg\left(  t\right)  \sum_{k=0}^{\infty}\frac{\left(
-1\right)  ^{k}a^{k}}{k!}\Phi^{bk+b-1}\left(  v\right)  \left[  1-\Phi\left(
v\right)  \right]  ^{-\left(  bk+b+1\right)  }.
\]
Since $0<\Phi\left(  v\right)  <1$, for $t>0$ and $\left(  bk+b+1\right)  >0$,
then by using the binomial series expansion $\left[  1-\Phi\left(  v\right)
\right]  ^{-\left(  bk+b+1\right)  }$ given by%
\[
\left[  1-\Phi\left(  v\right)  \right]  ^{-\left(  bk+b+1\right)  }%
=\sum_{j=0}^{\infty}\frac{\Gamma\left(  bk+b+1+j\right)  }{j!\Gamma\left(
bk+b+1\right)  }\Phi^{j}\left(  v\right)  ,
\]
where $\Gamma\left(  \cdot\right)  $ is the complete gamma function, we obtain%
\begin{equation}
f\left(  t\right)  =\sum_{k,j=0}^{\infty}w_{k,j}h_{bk+b+j}\left(  t\right)  ,
\label{9}%
\end{equation}
where%
\[
w_{k,j}=\frac{\left(  -1\right)  ^{k}ba^{k+1}\Gamma\left(  bk+b+1+j\right)
}{k!j!\left(  bk+b+j\right)  \Gamma\left(  bk+b+1\right)  },
\]
and $h_{bk+b+j}\left(  t\right)  \ $denotes the EBS$\left(  \alpha
,\beta,bk+b+j\right)  $\textbf{\ }density function. By integrating (\ref{9}),
we get%
\begin{equation}
F\left(  t\right)  =\sum_{k,j=0}^{\infty}w_{k,j}\Phi^{bk+b+j}\left(  v\right)
. \label{10}%
\end{equation}
\noindent It easy clear that $\sum_{k=0}^{\infty}\sum_{j=0}^{\infty}%
w_{k,j}=1.$ Equation (\ref{9}) means that the pdf of the WBS distribution is a
double linear mixture of the pdf of EBS distribution. Based on this equation,
several structural properties of the WBS distribution can be obtained by
knowing those of the EBS distribution. For example, the ordinary, inverse and
factorial moments, generating function and characteristic function of the WBS
distribution can be obtained directly from the\ EBS distribution.\medskip

\noindent If $b$ is a positive real non-integer, we can expand $\Phi
^{bk+b+j}\left(  v\right)  $ as%
\begin{equation}
\Phi^{bk+b+j}\left(  v\right)  =\sum_{r=0}^{\infty}s_{r}\left(  bk+b+j\right)
\Phi^{r}\left(  v\right)  , \label{11}%
\end{equation}
where%
\[
s_{r}\left(  m\right)  =\sum_{l=r}^{\infty}\left(  -1\right)  ^{l+r}\left(
\begin{tabular}
[c]{l}%
$m$\\
$l$%
\end{tabular}
\right)  \left(
\begin{tabular}
[c]{l}%
$l$\\
$r$%
\end{tabular}
\right)  .
\]
Thus, from equations (\ref{2}), (\ref{9}) and (\ref{11}), we get%
\begin{equation}
f\left(  t\right)  =g\left(  t\right)  \sum_{r=0}^{\infty}d_{r}\Phi^{r}\left(
v\right)  , \label{12}%
\end{equation}
where%
\[
d_{r}=\sum_{k,j=0}^{\infty}w_{k,j}s_{r}\left(  bk+b+j\right)  .
\]

\section{\textbf{Some structural properties\label{sec4}}}

\noindent In this section, we give some mathematical properties of the WBS distribution.

\subsection{Moments}

In this subsection, we derive the expression for $s$th order moment of WBS
distribution. The moments of some orders will help in determining the expected
life time of adevice and also the dispersion, skewness and kurtosis in a given
set of observations arising in reliability applications. The $s$th moment of
the WBS random variable $T$ can be derived from the probability weighted
moments of the BS distribution. The probability weighted moments of the BS
distribution are formally defined, for $p$ and $r$ non-negative integers, by%
\begin{equation}
\tau_{p,r}=\int_{0}^{\infty}t^{p}g\left(  t\right)  \Phi^{r}\left(  v\right)
dt. \label{13}%
\end{equation}
There are many softwares such as MAPLE, MATLAB and R that can be used to
compute the integral (\ref{13}) numerically. From \cite{r6}, we have an
alternative representation to compute $\tau_{p,r}$ that is%
\begin{align}
\tau_{p,r}  &  =\frac{\beta^{p}}{2^{r}}\sum_{j=1}^{r}\binom{r}{j}\sum
_{k_{1},\ldots,k_{j}=0}^{\infty}A\left(  k_{1},\ldots,k_{j}\right) \nonumber\\
&  \times\sum_{m=0}^{2s_{j}+j}\left(  -1\right)  ^{m}\binom{2s_{j}+j}%
{m}I\left(  p+\frac{2s_{j}+j-2m}{2},\alpha\right)  , \label{14}%
\end{align}
where $s_{j}=k_{1}+\ldots+k_{j},$ $A\left(  k_{1},\ldots,k_{j}\right)
=\alpha^{-2s_{j}-j}a_{k_{1}}\ldots a_{kj}$,

\noindent\ $a_{k}=\left(  -1\right)  ^{k}2^{\left(  1-2k\right)  /2}\left[
\sqrt{\pi}\left(  2k+1\right)  k!\right]  ^{-1}$ and $I\left(  p+\left(
2s_{j}+j-2m\right)  /2,\alpha\right)  $ is calculated from (\ref{33}) in terms
of the modified Bessel function of the third kind.\medskip

\noindent Therefore, the $s$th moment of $T$ can be written from (\ref{12}) as%
\begin{equation}
\mu_{s}^{\prime}=\sum_{r=0}^{\infty}d_{r}\tau_{s,r}, \label{144}%
\end{equation}
where $\tau_{s,r}$ is obtained from (\ref{14}) and $d_{r}$ is given by
(\ref{12}). We can compute numerically the $s$th moment in any symbolic
software by taking in the sum a large number of summands instead of infinity.

\subsection{Moment generating function}

Let $T$ $\sim$ WBS$\left(  \alpha,\beta,a,b\right)  $. The moment generating
function of $T$, say $M\left(  s\right)  =E\left(  e^{sT}\right)  $, is an
alternative specification of its probability distribution. Here, we provide
two representations for $M\left(  s\right)  $. From (\ref{12}), we obtain a
first representation%
\[
M\left(  s\right)  =\sum_{r=0}^{\infty}d_{r}\int_{0}^{\infty}e^{st}g\left(
t\right)  \Phi^{r}\left(  v\right)  dt.
\]
Expanding the exponential function, we can rewritte $M\left(  s\right)  $ as%
\[
M\left(  s\right)  =\sum_{r=0}^{\infty}\sum_{p=0}^{\infty}\frac{d_{r}%
\tau_{p,r}}{p!}s^{p}.
\]
The second representation for$M\left(  s\right)  $ is based on the quantile
expansion of the BS distribution. From (\ref{9}), we have%
\[
M\left(  s\right)  =\sum_{k,j=0}^{\infty}w_{k,j}\int_{0}^{\infty}%
e^{st}g\left(  t\right)  \Phi^{bk+b+j-1}\left(  v\right)  dt,
\]
where $g\left(  t\right)  $ is the BS($\alpha$,$\beta$) pdf. By setting
$u=\Phi\left(  v\right)  $ in above integral, we get%
\[
M\left(  s\right)  =\sum_{k,j=0}^{\infty}w_{k,j}\int_{0}^{1}u^{bk+b+j-1}%
\exp\left(  sQ\left(  u\right)  \right)  du,
\]
where $t=Q(u)$ is the quantile function of the BS distribution and
$u=\Phi\left(  v\right)  $ is given by (\ref{1}). Using the exponential
expansion, we get%
\begin{equation}
M\left(  s\right)  =\sum_{k,j=0}^{\infty}w_{k,j}\sum_{p=0}^{\infty}\frac
{s^{p}}{p!}\int_{0}^{1}u^{bk+b+j-1}Q^{p}\left(  u\right)  du. \label{15}%
\end{equation}
From \cite{r6}, if the condition $-2<\left(  t/\beta\right)  ^{1/2}-\left(
t/\beta\right)  ^{-1/2}<2$ holds, we have the expansion for the quantile
function of the BS distribution
\begin{equation}
t=Q\left(  u\right)  =\sum_{i=0}^{\infty}\eta_{i}\left(  u-1/2\right)  ^{i},
\label{16}%
\end{equation}
where%
\[
\eta_{i}=\left(  2\pi\right)  ^{i/2}\sum_{j=0}^{\infty}d_{j}e_{j,i},
\]
$d_{0}=\beta,$ $d_{2j+1}=\beta\alpha^{2j+1}\binom{1/2}{j}2^{-2j}$ for
$j\geq0,$ $d_{2}=\beta\alpha^{2}/2,$ $d_{2j}=0$ for $j\geq2\ $and the
quantities $e_{j,i}$ can be determined from the recurrence equation%
\[
e_{j,i}=\left(  ia_{0}\right)  ^{-1}\sum_{m=1}^{i}\left(  mj+m-i\right)
a_{m}c_{j,i-m},
\]
and $e_{j,0}=a_{0}^{j}$ . Here, the quantities $a_{m}$ are defined by
$a_{m}=0$ (for $m=0,2,4,\ldots$) and $a_{m}=b_{(m-1)}/2$ (for $i=1,3,5,\ldots
$), where the $b_{m}$'s are computed recursively from%
\[
b_{m+1}=\frac{1}{2\left(  2m+3\right)  }\sum_{r=1}^{m}\frac{\left(
2r+1\right)  \left(  2m-2r+1\right)  b_{r}b_{m-r}}{\left(  r+1\right)  \left(
2r+1\right)  }.
\]
The first constants are We have $b_{0}=1,$ $b_{1}=1/6,$ $b_{2}=7/120,$
$b_{3}=127/7560,...$.\bigskip

\noindent Inserting (\ref{16}) in (\ref{15}), we get%
\begin{equation}
M\left(  s\right)  =\sum_{k,j=0}^{\infty}w_{k,j}\sum_{p=0}^{\infty}\frac
{s^{p}}{p!}\int_{0}^{1}u^{bk+b+j-1}\left(  \sum_{i=0}^{\infty}\eta_{i}\left(
u-1/2\right)  ^{i}\right)  ^{p}du. \label{17}%
\end{equation}
From \cite[Sec. 0.314]{r12} for a power series raised to a positive integer
$p,$ we have
\[
\left(  \sum_{i=0}^{\infty}\eta_{i}\left(  u-1/2\right)  ^{i}\right)
^{p}=\sum_{i=0}^{\infty}\delta_{p,i}\left(  u-1/2\right)  ^{i},
\]
where the coefficients $\delta_{p,i}$ (for $i=1,2,...$) can be determined from
the recurrence equation%
\[
\delta_{p,i}=\left(  ia_{0}\right)  ^{-1}\sum_{m=1}^{i}\left(  mp+m-i\right)
a_{m}\delta_{j,i-m},
\]
and $\delta_{p,0}=a_{0}^{j}$ . Hence, $\delta_{j,i}$ comes directly from
$\delta_{j,0},\ldots,\delta_{j,i-1}$ and, therefore, from $a_{0},...,a_{i}$.
Then%
\begin{equation}
M\left(  s\right)  =\sum_{k,j,p,i=0}^{\infty}w_{k,j}\frac{s^{p}}{p!}%
\delta_{p,i}\int_{0}^{1}u^{b\left(  k+1\right)  +j-1}\left(  u-1/2\right)
^{i}du. \label{18}%
\end{equation}
Therefore, using the binomial expansion in (\ref{18}), we obtain%
\[
M\left(  s\right)  =\sum_{k,j,i,p=0}^{\infty}\sum_{l=0}^{i}\frac{\left(
-1\right)  ^{i-l}\binom{i}{l}w_{k,j}\delta_{p,i}}{p!\left(  bk+b+j+l\right)
2^{i-l}}s^{p}.
\]

\subsection{Quantile function and simulation}

In this subsection, we give an expression for WBS quantile function, $Q\left(
u\right)  =F^{-1}(u),$ in terms of the BS quantile function $Q_{BS}\left(
\cdot\right)  $ . The BS quantile function is straightforward computed from
the standard normal quantile function $\Phi^{-1}\left(  u\right)  $. We have
(see , \cite{r6})%
\[
Q_{BS}\left(  u\right)  =\frac{\beta}{2}\left(  \alpha\Phi^{-1}\left(
u\right)  +\sqrt{4+\left[  \alpha\Phi^{-1}\left(  u\right)  \right]  ^{2}%
}\right)  ^{2}.
\]
Then, by inverting $F(x)=u$, we obtain%
\[
Q_{WBS}\left(  u\right)  =\frac{\beta}{2}\left(  \alpha\Phi^{-1}\left(
p\right)  +\sqrt{4+\left[  \alpha\Phi^{-1}\left(  p\right)  \right]  ^{2}%
}\right)  ^{2},
\]
where%
\[
p=\frac{\left(  -\frac{1}{a}\ln\left(  1-u\right)  \right)  ^{\frac{1}{b}}%
}{1+\left(  -\frac{1}{a}\ln\left(  1-u\right)  \right)  ^{\frac{1}{b}}}.
\]
\noindent Therefore, it is easy to simulate the WBS distribution. Let $U$ be a
continuous uniform variable on the unit interval $\left(  0,1\right]  $. Thus,
using the inverse transformation method, the random variable $T$ given by%
\begin{equation}
T=Q_{WBS}(U)=\frac{\beta}{2}\left(  \alpha\Phi^{-1}\left(  P\right)
+\sqrt{4+\left[  \alpha\Phi^{-1}\left(  P\right)  \right]  ^{2}}\right)  ^{2},
\label{19}%
\end{equation}
where%
\[
P=\frac{\left(  -\frac{1}{a}\ln\left(  1-U\right)  \right)  ^{\frac{1}{b}}%
}{1+\left(  -\frac{1}{a}\ln\left(  1-U\right)  \right)  ^{\frac{1}{b}}},
\]
has the WBS distribution. Equation (\ref{19}) may be used to generate random
numbers from the WBS distribution when the parameters are known.\medskip

\noindent We plot the exact and the empirical cdf of WBS distribution in
Figure 3 using a pseudo random sample of size1000 to check the correctness of
the procedure for simulating a data set from WBS distribution. The histograms
for two generated data sets and the exact WBS density plots from two simulated
data sets for some parameter values are given in Figure 4. These plots
indicate that the simulated values are consistent with the WBS distribution.%

%TCIMACRO{\FRAME{itbpFU}{4.8922in}{2.6602in}{0in}{\Qcb{Figure 3. Comparison of
%exact and empirical cdf of the WBS distribution to simulate random numbers.}%
%}{}{fig3.eps}{\special{ language "Scientific Word";  type "GRAPHIC";
%display "USEDEF";  valid_file "F";  width 4.8922in;  height 2.6602in;
%depth 0in;  original-width 18.9204in;  original-height 9.1584in;
%cropleft "0";  croptop "0.9150";  cropright "0.9969";  cropbottom "0.0020";
%filename 'fig3.eps';file-properties "XNPEU";}} }%
%BeginExpansion
{\parbox[b]{4.8922in}{\begin{center}
\ifcase\msipdfoutput
\includegraphics[
trim=0.000000in 0.018317in 0.058653in 0.778464in,
height=2.6602in,
width=4.8922in
]%
{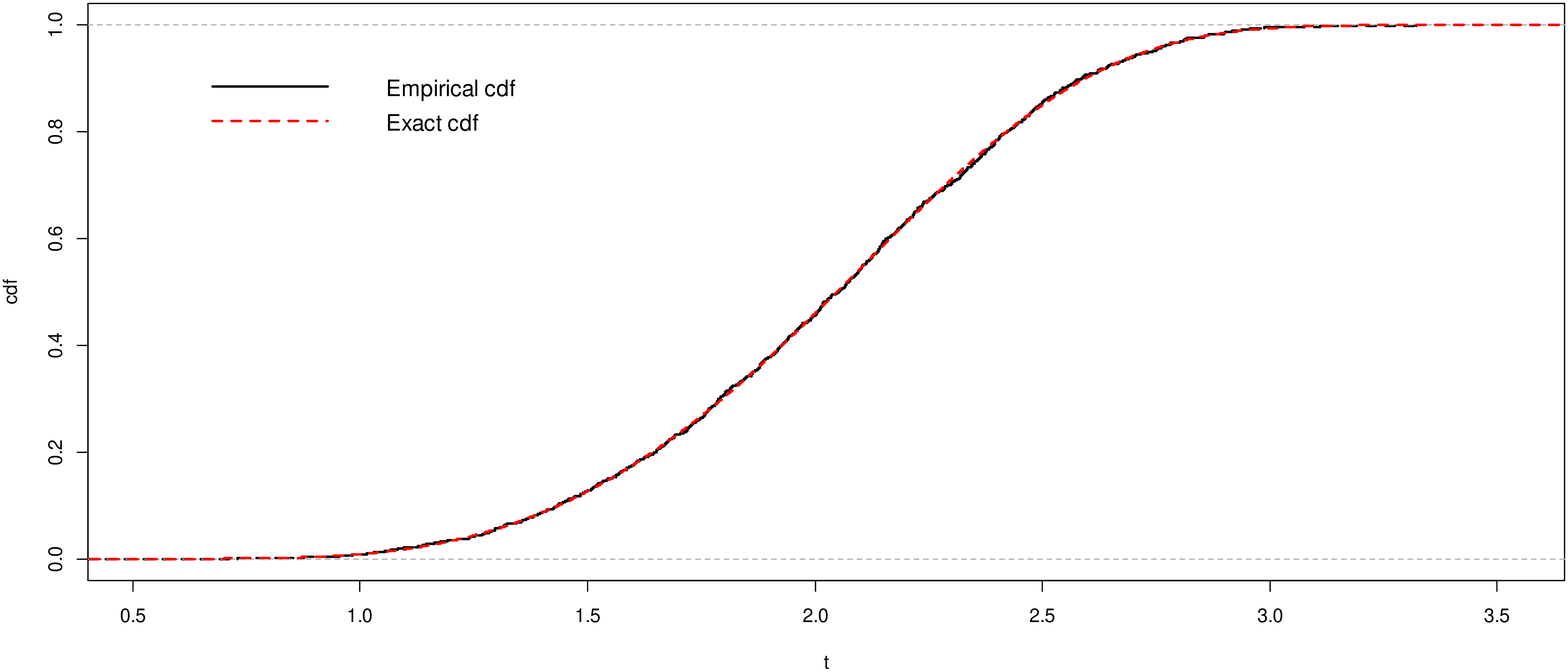}%
\else
\includegraphics[
height=2.6602in,
width=4.8922in
]%
{C:/Users/LAZHAR/IDEES/Weibull-Birnbaum-Saunders distribution/arXiv/graphics/fig3.jpg}%
\fi
\\
Figure 3. Comparison of exact and empirical cdf of the WBS distribution to
simulate random numbers.
\end{center}}}
%EndExpansion

\bigskip%
%TCIMACRO{\FRAME{itbpFU}{4.951in}{2.5918in}{0in}{\Qcb{Figure 4. Plots of the
%WBS densities for simulated data sets: (a) $\alpha=2.5,$ $\beta=2.5$, $a=2$
%and $b=4$; (b) $\alpha=0.5,$ $\beta=1.7$, $a=0.2$ and $b=0.4$.}}{}%
%{fig4.eps}{\special{ language "Scientific Word";  type "GRAPHIC";
%display "USEDEF";  valid_file "F";  width 4.951in;  height 2.5918in;
%depth 0in;  original-width 18.9204in;  original-height 9.1584in;
%cropleft "0";  croptop "1";  cropright "1";  cropbottom "0";
%filename 'fig4.eps';file-properties "XNPEU";}} }%
%BeginExpansion
{\parbox[b]{4.951in}{\begin{center}
\ifcase\msipdfoutput
\includegraphics[
height=2.5918in,
width=4.951in
]%
{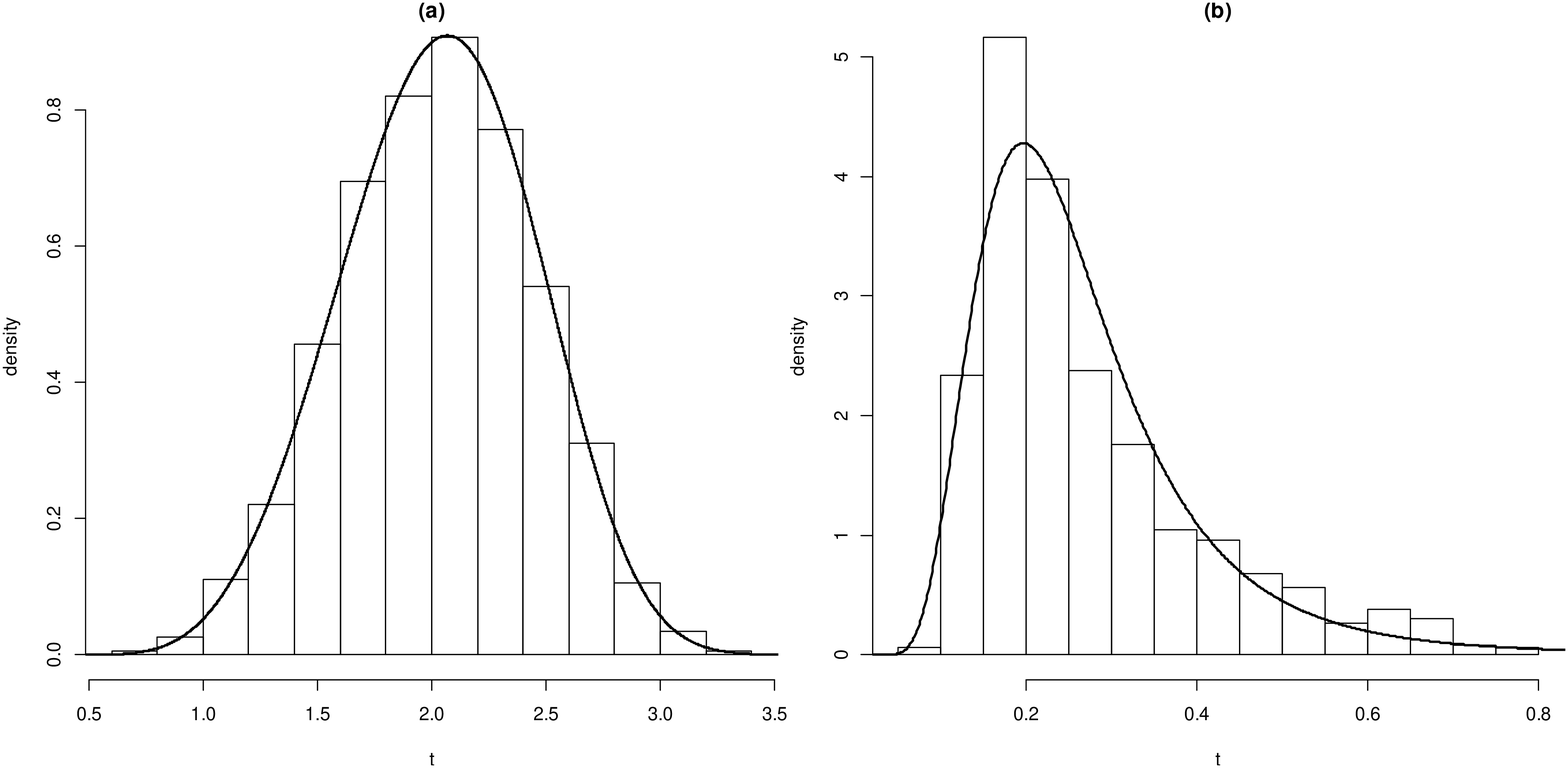}%
\else
\includegraphics[
height=2.5918in,
width=4.951in
]%
{C:/Users/LAZHAR/IDEES/Weibull-Birnbaum-Saunders distribution/arXiv/graphics/fig4.jpg}%
\fi
\\
Figure 4. Plots of the WBS densities for simulated data sets: (a)
$\alpha=2.5,$ $\beta=2.5$, $a=2$ and $b=4$; (b) $\alpha=0.5,$ $\beta=1.7$,
$a=0.2$ and $b=0.4$.
\end{center}}}
%EndExpansion

\bigskip

\subsection{Mean deviations}

Let $T$ be a random variable having the WBS($\alpha,\beta,a,b$) distribution.
The mean deviations of $T$ about the mean and about the median can be used as
measures of spread in a population. They are given by%
\[
\delta_{1}=E\left(  \left\vert T-\mu_{1}^{\prime}\right\vert \right)
=\int_{0}^{\infty}\left\vert T-\mu_{1}^{\prime}\right\vert f\left(  t\right)
dt
\]
and%
\[
\delta_{2}=E\left(  \left\vert T-m\right\vert \right)  =\int_{0}^{\infty
}\left\vert t-m\right\vert f\left(  t\right)  dt,
\]
\ \ respectively, where the mean $\mu_{1}^{\prime}$ is calculated from
(\ref{144}) and the median $m$ is given by $m=Q_{WBS}(1/2).$ The measures
$\delta_{1}$ and \ $\delta_{2}$ can be expressed as%
\[
\delta_{1}=2\mu_{1}^{\prime}F\left(  \mu_{1}^{\prime}\right)  -J\left(
\mu_{1}^{\prime}\right)  \text{ \ \ \ and \ \ \ \ }\delta_{2}=E\left(
\left\vert T-m\right\vert \right)  =\mu_{1}^{\prime}-2J\left(  m\right)  ,
\]
where $J\left(  q\right)  =\int_{0}^{q}tf\left(  t\right)  dt$.\ From
(\ref{12}), $J\left(  q\right)  $ can be written as%
\begin{equation}
J\left(  q\right)  =\sum_{r=0}^{\infty}d_{r}\varphi\left(  q,r\right)  ,
\label{20}%
\end{equation}
where%
\[
\varphi\left(  q,r\right)  =\int_{0}^{q}tg\left(  t\right)  \Phi^{r}\left(
v\right)  dt.
\]
From \cite{r6}, we have
\begin{align*}
\varphi\left(  q,r\right)   &  =\frac{\kappa\left(  \alpha,\beta\right)
}{2^{r}}\sum_{j=1}^{r}\binom{r}{j}\sum_{k_{1},\ldots,k_{j}=0}^{\infty}%
\beta^{-\left(  2s_{j}+j\right)  /2}A\left(  k_{1},\ldots,k_{j}\right)
\sum_{m=0}^{2s_{j}+j}\left(  -\beta\right)  ^{m}\\
&  \times\binom{2s_{j}+j}{m}\int_{0}^{q}t^{\left(  2s_{j}+j-2m-1\right)
/2}\left(  t+\beta\right)  \exp\left\{  -\frac{\tau\left(  t/\beta\right)
}{2\alpha^{2}}\right\}  dt,
\end{align*}
where $s_{j}$ and $A\left(  k_{1},\ldots,k_{j}\right)  $ are defined in
(\ref{14}). Consider%
\[
D\left(  p,q\right)  =\int_{0}^{q}t^{p}\exp\left\{  -\frac{\left(
t/\beta+\beta/t\right)  }{2\alpha^{2}}\right\}  dt=\beta^{p+1}\int%
_{0}^{q/\beta}u^{p}\exp\left\{  -\frac{u+u^{-1}}{2\alpha^{2}}\right\}  du.
\]
From \cite{r26}, we can write%
\[
D\left(  p,q\right)  =2\beta^{p+1}K_{p+1}\left(  \alpha^{-2}\right)
-q^{p+1}K_{p+1}\left(  \frac{q}{2\alpha^{2}\beta},\frac{\beta}{2\alpha^{2}%
q}\right)  ,
\]
where, $K_{v}\left(  z_{1},z_{2}\right)  $ is the incomplete Bessel function
with order $v$ and arguments $z_{1}$ and $z_{2}$. Then, we obtain%
\begin{align*}
\varphi\left(  q,r\right)   &  =\frac{\kappa\left(  \alpha,\beta\right)
}{2^{r}}\sum_{j=1}^{r}\binom{r}{j}\sum_{k_{1},\ldots,k_{j}=0}^{\infty}%
\beta^{-\left(  2s_{j}+j\right)  /2}A\left(  k_{1},\ldots,k_{j}\right)
\sum_{m=0}^{2s_{j}+j}\left(  -\beta\right)  ^{m}\\
&  \times\binom{2s_{j}+j}{m}\left\{  D\left(  \frac{2s_{j}+j-2m+1}%
{2},q\right)  +\beta D\left(  \frac{2s_{j}+j-2m-1}{2},q\right)  \right\}  ,
\end{align*}
which can be calculated from the function $D\left(  p,q\right)  $. Hence, we
can use this expression for $\varphi\left(  q,r\right)  $ to compute $J\left(
q\right)  $. From (\ref{20}), we obtain the Bonferroni and Lorenz curves
defined by $B\left(  p\right)  =J\left(  q\right)  /p\mu_{1}^{\prime}\ $and
$L\left(  t\right)  =J\left(  q\right)  /\mu_{1}^{\prime}$, respectively.
These curves have applications in reliability.\bigskip\ 

\subsection{Reliability}

In the stress-strength modelling, $R=\mathbb{P}(T_{2}<T_{1})$ is a measure of
component reliability when it is subjected to random stress $T_{2}$ and has
strength $T_{1}$. The component fails at the instant that the stress applied
to it exceeds the strength and the component will function satisfactorily
whenever $T_{1}>T_{2}$. The parameter $R$ is referred to as the reliability
parameter. This type of functional can be of practical importance in many
applications. In this Section, we derive the reliability $R$ when $T_{1}$ and
$T_{2}$ have independent WBS$\left(  \alpha,\beta,a_{1},b_{1}\right)  $ and
WBS$\left(  \alpha,\beta,a_{2},b_{2}\right)  $ distributions. The pdf of
$T_{1}$ and the cdf of $T_{2}$ can be obtained from (\ref{9}) and (\ref{10})
as%
\[
f_{1}\left(  t\right)  =g\left(  t\right)  \sum_{k,j=0}^{\infty}w_{1k,j}%
^{\ast}\Phi^{b_{1}k+b_{1}+j-1},
\]
and%
\[
F_{2}\left(  t\right)  =\sum_{l,m=0}^{\infty}w_{2l,m}\Phi^{b_{2}l+b_{2}%
+m}\left(  v\right)
\]
respectively, where%
\[
w_{1k,j}^{\ast}=\frac{\left(  -1\right)  ^{k}b_{1}a_{1}^{k+1}\Gamma\left(
b_{1}k+b_{1}+1+j\right)  }{k!j!\Gamma\left(  b_{1}k+b_{1}+1\right)  }.
\]
and%
\[
w_{2l,m}=\frac{\left(  -1\right)  ^{l}b_{2}a_{2}^{l+1}\Gamma\left(
b_{2}l+b_{2}+1+m\right)  }{l!m!\left(  b_{2}l+b_{2}+m\right)  \Gamma\left(
b_{2}l+b_{2}+1\right)  }.
\]
We have%
\[
R=\int_{0}^{\infty}f_{1}\left(  t\right)  F_{2}\left(  t\right)  dt.
\]
Then%
\[
R=\sum_{k,j,l,m=0}^{\infty}w_{1k,j}^{\ast}w_{2l,m}\int_{0}^{\infty}g\left(
t\right)  \Phi^{d^{\ast}}\left(  v\right)  dt,
\]
where%
\[
d^{\ast}=b_{1}\left(  k+1\right)  +b_{2}\left(  l+1\right)  +j+m-1.
\]
From (\ref{11}), we can write%
\[
\Phi^{d^{\ast}}\left(  v\right)  =\sum_{r=0}^{\infty}s_{r}\left(
\delta\right)  \Phi^{r}\left(  v\right)  ,
\]
and then, we get%
\[
R=\sum_{k,j,l,m=0}^{\infty}w_{1k,j}^{\ast}w_{2l,m}\sum_{r=0}^{\infty}%
s_{r}\left(  \delta\right)  \tau_{0,r},
\]
where $\tau_{0,r}$ is obtained from (\ref{14}).

\subsection{Order statistics}

In this section, the distribution of the $i$th order statistic for the WBS
distribution are presented. The order statistics play an important role in
reliability and life testing. Let $T_{1},\ldots,T_{n}$ be a simple random
sample from WBS distribution with cdf and pdf as in (\ref{6}) and (\ref{7}),
respectively. Let $T_{1,n}\leq\cdots\leq T_{n,n}$ denote the order statistics
obtained from this sample. In reliability literature, the $i$th order
statistic, say $T_{i;n}$, denotes the lifetime of an $(n-i+1)$--out--of--$n$
system which consists of $n$ independent and identically components. \ The pdf
of $T_{i;n}$ is given by%
\begin{equation}
f_{i,n}\left(  t\right)  =\frac{n!}{\left(  n-i\right)  !\left(  n-1\right)
!}\sum_{j=0}^{n-i}\left(  -1\right)  ^{j}\binom{n-i}{j}f\left(  t\right)
F\left(  t\right)  ^{^{i+j-1}}. \label{21}%
\end{equation}
\bigskip

\noindent From (\ref{6}), we have%
\[
F^{i+j-1}\left(  t\right)  =\left[  1-\exp\left(  -a\left[  \frac{\Phi\left(
v\right)  }{1-\Phi\left(  v\right)  }\right]  ^{b}\right)  \right]  ^{i+j-1}.
\]
Using the binomial series expansion, we get%
\begin{equation}
F^{i+j-1}\left(  t\right)  =\sum_{k=0}^{\infty}\left(  -1\right)  ^{k}%
\binom{i+j-1}{k}\exp\left(  -ka\left[  \frac{\Phi\left(  v\right)  }%
{1-\Phi\left(  v\right)  }\right]  ^{b}\right)  . \label{22}%
\end{equation}
Inserting (\ref{7}) and (\ref{22}) in (\ref{21}), we obtain%
\begin{align*}
f_{i,n}\left(  t\right)   &  =\frac{n!ab\kappa\left(  \alpha,\beta\right)
t^{-3/2}\left(  t+\beta\right)  }{\left(  n-i\right)  !\left(  n-1\right)
!}\exp\left\{  -\frac{\tau\left(  t/\beta\right)  }{2\alpha^{2}}\right\}
\left[  \frac{\Phi\left(  v\right)  ^{b-1}}{\left\{  1-\Phi\left(  v\right)
\right\}  ^{b+1}}\right] \\
&  \times\sum_{j=0}^{n-k}\sum_{k=0}^{\infty}\left(  -1\right)  ^{k+j}%
\binom{i+j-1}{k}\binom{n-i}{j}\exp\left(  -a\left(  k+1\right)  \left[
\frac{\Phi\left(  v\right)  }{1-\Phi\left(  v\right)  }\right]  ^{b}\right)  .
\end{align*}
Using the power series for the exponential function, we have%
\[
\exp\left(  -a\left(  k+1\right)  \left[  \frac{\Phi\left(  v\right)  }%
{1-\Phi\left(  v\right)  }\right]  ^{b}\right)  =\sum_{l=0}^{\infty}%
\frac{\left(  -1\right)  ^{l}\left(  ak+a\right)  ^{l}}{l!}\frac{\Phi\left(
v\right)  ^{bl}}{\left[  1-\Phi\left(  v\right)  \right]  ^{bl}}.
\]
Then%
\begin{align*}
f_{i,n}\left(  t\right)   &  =\frac{g(t)n!ab}{\left(  n-i\right)  !\left(
n-1\right)  !}\sum_{j=0}^{n-k}\sum_{k=0}^{\infty}\left(  -1\right)
^{k+j}\binom{i+j-1}{k}\binom{n-i}{j}\\
&  \times\sum_{l=0}^{\infty}\frac{\left(  -1\right)  ^{l}\left(  ak+a\right)
^{l}}{l!}\Phi\left(  v\right)  ^{bl+b-1}\left[  1-\Phi\left(  v\right)
\right]  ^{-\left(  bl+b+1\right)  }.
\end{align*}
Since $0<\Phi\left(  v\right)  <1$, for $t>0$ and $\left(  bl+b+1\right)  >0$,
we have%
\[
\left[  1-\Phi\left(  v\right)  \right]  ^{-\left(  bl+b+1\right)  }%
=\sum_{m=0}^{\infty}\frac{\Gamma\left(  bl+b+1+m\right)  }{m!\Gamma\left(
bl+b+1\right)  }\Phi^{m}\left(  v\right)  .
\]
Therefore, the pdf of the $i$th order statistic for WBS distribution is%
\begin{equation}
f_{i,n}\left(  t\right)  =\sum_{l,m=0}^{\infty}\vartheta_{l,m}h_{bl+b+m}%
\left(  t\right)  , \label{23}%
\end{equation}
where%
\[
\vartheta_{l,m}=\sum_{k=0}^{\infty}\sum_{j=0}^{n-i}\frac{\left(  -1\right)
^{k+j+l}n!ba^{k+1}\left(  k+1\right)  ^{l}\Gamma\left(  bl+b+1+m\right)
}{l!m!\left(  n-i\right)  !\left(  n-1\right)  !\left(  bl+b+m\right)
\Gamma\left(  bl+b+1\right)  }\binom{i+j-1}{k}\binom{n-i}{j},
\]
and $h_{bl+b+m}\ $ is the EBS density function with power parameter $bl+b+m$.
Equation (\ref{23}) means that the density function of the WBS order
statistics is a linear mixture of the EBS densities. Then, we can easily
obtain the mathematical properties for $T_{i,n}$. For example, the $p$th
moment of $T_{i,n}$ is%
\[
E\left(  T_{i,n}^{p}\right)  =\sum_{l,m=0}^{\infty}\vartheta_{l,m}\left(
bl+l+m\right)  \tau_{p,\left(  bl+l+m-1\right)  }.
\]
\bigskip

\section{\textbf{Estimation\label{sec5}}}

In this section, we consider estimation of the unknown parameters of the WBS
distribution by the method of maximum likelihood. Let $x_{1},x_{2}%
,\ldots,x_{n}$ be observed values of $X_{1},X_{2},\ldots,X_{n}$, $n$
independent random variables having the WBS distribution with unknown
parameter vector $\mathbf{\xi}=(\alpha,\beta,a,b)^{T}$. The total
log-likelihood function for $\mathbf{\xi}$, is given by%
\begin{align*}
\ell &  =\ell\left(  \mathbf{\xi}\right)  =n\log\left(  a\right)  +\log\left(
b\right)  +\log\left[  \kappa\left(  \alpha,\beta\right)  \right]  -\frac
{3}{2}\sum_{i=1}^{n}\log\left(  t_{i}\right)  +\sum_{i=1}^{n}\log\left(
t_{i}+\beta\right) \\
&  -\frac{1}{2\alpha^{2}}\sum_{i=1}^{n}\tau\left(  \frac{t_{i}}{\beta}\right)
+\left(  b-1\right)  \sum_{i=1}^{n}\log\left[  \Phi\left(  v_{i}\right)
\right]  -\left(  b+1\right)  \sum_{i=1}^{n}\log\left[  1-\Phi\left(
v_{i}\right)  \right] \\
&  -a\sum_{i=1}^{n}\left[  \frac{\Phi\left(  v_{i}\right)  }{1-\Phi\left(
v_{i}\right)  }\right]  ^{b}.
\end{align*}
Then, the components of the unit score vector $\mathbf{U=U}\left(
\mathbf{\xi}\right)  =\left(  \partial\ell/\partial\alpha,\partial
\ell/\partial\beta,\partial\ell/\partial a,\partial\ell/\partial b\right)
^{T}$ are given by%
\begin{align*}
\frac{\partial\ell}{\partial\alpha}  &  =-\frac{n}{\alpha}\left(  1+\frac
{2}{\alpha^{2}}\right)  +\frac{1}{\alpha^{3}}\sum_{i=1}^{n}\left(  \frac
{t_{i}}{\beta}+\frac{\beta}{t_{i}}\right)  -\frac{\left(  b-1\right)  }%
{\alpha}\sum_{i=1}^{n}\frac{v_{i}\phi\left(  v_{i}\right)  }{\Phi\left(
v_{i}\right)  }\\
&  -\frac{\left(  b+1\right)  }{\alpha}\sum_{i=1}^{n}\frac{v_{i}\phi\left(
v_{i}\right)  }{1-\Phi\left(  v_{i}\right)  }+\frac{ab}{\alpha}\sum_{i=1}%
^{n}\frac{v_{i}\phi\left(  v_{i}\right)  \Phi\left(  v_{i}\right)  ^{b-1}%
}{\left[  1-\Phi\left(  v_{i}\right)  \right]  ^{b+1}},\\
\frac{\partial\ell}{\partial\beta}  &  =-\frac{n}{2\beta}+\sum_{i=1}^{n}%
\frac{1}{t_{i}+\beta}+\frac{1}{2\beta\alpha^{2}}\sum_{i=1}^{n}\left(
\frac{t_{i}}{\beta}-\frac{\beta}{t_{i}}\right)  -\frac{\left(  b-1\right)
}{2\beta\alpha}\sum_{i=1}^{n}\frac{\tau\left(  \sqrt{t_{i}/\beta}\right)
\phi\left(  v_{i}\right)  }{\Phi\left(  v_{i}\right)  }\\
&  -\frac{\left(  b+1\right)  }{2\beta\alpha}\sum_{i=1}^{n}\frac{\tau\left(
\sqrt{t_{i}/\beta}\right)  \phi\left(  v_{i}\right)  }{1-\Phi\left(
v_{i}\right)  }+\frac{ab}{2\beta\alpha}\sum_{i=1}^{n}\frac{\tau\left(
\sqrt{t_{i}/\beta}\right)  \phi\left(  v_{i}\right)  \Phi\left(  v_{i}\right)
^{b-1}}{\left[  1-\Phi\left(  v_{i}\right)  \right]  ^{b+1}},\\[0.2cm]
\frac{\partial\ell}{\partial a}  &  =\frac{n}{a}-\sum_{i=1}^{n}\left[
\frac{\Phi\left(  v_{i}\right)  }{1-\Phi\left(  v_{i}\right)  }\right]  ^{b},
\end{align*}
and%
\begin{align*}
\frac{\partial\ell}{\partial b}  &  =\frac{n}{b}+\sum_{i=1}^{n}\log\left[
\Phi\left(  v_{i}\right)  \right]  -\sum_{i=1}^{n}\log\left[  1-\Phi\left(
v_{i}\right)  \right] \\
&  -a\sum_{i=1}^{n}\left[  \frac{\Phi\left(  v_{i}\right)  }{1-\Phi\left(
v_{i}\right)  }\right]  ^{b}\log\left[  \frac{\Phi\left(  v_{i}\right)
}{1-\Phi\left(  v_{i}\right)  }\right]  ,
\end{align*}
where $\phi\left(  \cdot\right)  $ is the standard normal density function,
$\tau\left(  \sqrt{t_{i}/\beta}\right)  =\sqrt{t_{i}/\beta}+\sqrt{\beta/t_{i}%
}$ and $v_{i}=\alpha^{-1}\left\{  \sqrt{t_{i}/\beta}-\sqrt{\beta/t_{i}%
}\right\}  $ for $i=1,\ldots,n$. The maximum likelihood estimate
$\widehat{\mathbf{\xi}}$ of $\mathbf{\xi}\ $ is obtained by setting these
equations to zero, $\mathbf{U}\left(  \mathbf{\xi}\right)  =0,$\ solving them
simultaneously$.$ These equations cannot be solved analytically and
statistical software can be used to solve them numerically via iterative
methods such as the Newton--Raphson algorithm$.$\bigskip

\noindent We can use the normal approximation of the MLE of $\mathbf{\xi}$ to
construct approximate confidence intervals for the parameters. Under some
regular conditions (see \cite[Chapter 9]{r9}) that are fulfilled for the
parameters in the interior of the parameter space, the asymptotic distribution
of $\sqrt{n}\left(  \widehat{\mathbf{\xi}}-\mathbf{\xi}\right)  \ $is
multivariate normal $\mathcal{N}_{4}\left(  \mathbf{0},I^{-1}\left(
\mathbf{\xi}\right)  \right)  ,$ where $I\left(  \mathbf{\xi}\right)  $ is the
expected information matrix.$\ $This asymptotic behavior is valid if $I\left(
\mathbf{\xi}\right)  \ $is replaced by the observed information matrix,
$J\left(  \mathbf{\xi}\right)  $, evaluated at $\widehat{\mathbf{\xi}},$i.e.
$J\left(  \widehat{\mathbf{\xi}}\right)  .$ The observed information matrix is
given by%
\[
J\left(  \mathbf{\xi}\right)  =-\left(
\begin{array}
[c]{cccc}%
L_{\alpha\alpha} & L_{\alpha\beta} & L_{\alpha a} & L_{\alpha b}\\
. & L_{\beta\beta} & L_{\beta a} & I_{\beta b}\\
. & . & I_{aa} & I_{ab}\\
. & . & . & I_{bb}%
\end{array}
\right)  ,
\]
whose elements are given in the Appendix. The approximate $100(1-\eta)\%$
two-sided confidence intervals for $\alpha,\beta,a$ and $b$ are given by
$\widehat{\alpha}\pm z_{\frac{\eta}{2}}\sqrt{var\left(  \widehat{\alpha
}\right)  },$ $\widehat{\beta}\pm z_{\frac{\eta}{2}}\sqrt{var\left(
\widehat{\beta}\right)  },$ $\widehat{a}\pm z_{\frac{\eta}{2}}\sqrt{var\left(
\widehat{a}\right)  }\ $and $\widehat{b}\pm z_{\frac{\eta}{2}}\sqrt{var\left(
\widehat{b}\right)  }\ $respectively, where $z_{\frac{\eta}{2}}$ is the
quantile $\left(  1-\frac{\eta}{2}\right)  $ of the standard normal
distribution and $var\left(  \cdot\right)  $ is the diagonal element of
$J^{-1}\left(  \widehat{\mathbf{\xi}}\right)  $ corresponding to each
parameter.\bigskip

\section{\textbf{Applications\label{sec6}}}

In this section, we demonstrate the applicability and flexibility of the WBS
distribution by means of two well-known real data sets with different
shapes\ \ The first data set is given by Meeker and Escobar data \cite{r18}
and the second data set is given in \cite{r27}. The first data set has a
bathtub shaped failure rate function whereas the second data set has an
increasing failure rate function.\medskip

\noindent For these data sets, we compare the results of fitting the WBS
distribution with the Beta BS (BBS), Kumaraswamy BS (KBS), McDonald BS (McBS),
Marshall-Olkin extended BS (MOEBS), gamma BS (GBS), EBS and BS distributions
using the graphical method, minus twice the maximized log-likelihood
($-2\widehat{\ell}$), Akaike information criterion (AIC), Bayesian information
criterion (BIC), Consistent akaike information criterion\ (CAIC) and
Kolmogorov--Smirnov (K-S) test. The pdfs of the BBS, KBS, McBS, MOEBS and GBS
distributions (for $t>0$) are given by%
\[
f_{1}\left(  t\right)  =\frac{g\left(  t\right)  }{B\left(  a,b\right)  }%
\Phi^{a-1}\left(  v\right)  \left[  1-\Phi\left(  v\right)  \right]
^{b-1}\text{, }f_{2}\left(  t\right)  =abg\left(  t\right)  \Phi^{a-1}\left(
v\right)  \left[  1-\Phi^{a}\left(  v\right)  \right]  ^{b-1}\text{,}%
\]%
\[
f_{3}\left(  t\right)  =\frac{cg\left(  t\right)  }{B\left(  a/c,b\right)
}\Phi^{a-1}\left(  v\right)  \left[  1-\Phi^{c}\left(  v\right)  \right]
^{b-1},\text{ }f_{4}\left(  x\right)  =\frac{ag\left(  t\right)  }{\left[
1-\left(  1-a\right)  \Phi\left(  -v\right)  \right]  ^{2}}%
\]
and $f_{5}\left(  t\right)  =\dfrac{g\left(  t\right)  }{\Gamma\left(
a\right)  }\left[  -\log\left\{  1-\Phi\left(  v\right)  \right\}  \right]
^{b-1}\ $respectively, where $g\left(  t\right)  $ is the BS($\alpha$,$\beta$)
pdf (\ref{2}).and $\alpha,$ $\beta,$ $a,$ $b,$ $c>0$.\medskip

\subsection{Meeker and Escobar data}

The first data set represents failure and running times of 30 devices provided
by Meeker and Escobar \cite[ p.383]{r18}. The data set is: 2, 10, 13, 23, 23,
28, 30, 65 ,80, 88, 106, 143, 147, 173, 181, 212, 245, 247, 261, 266, 275,
293, 300, 300, 300, 300, 300, 300, 300, 300. The total time on test (TTT) plot
for the Meeker and Escobar data in Figure 5(a) shows a convex shape followed
by a concave shape. This corresponds to a bathtub-shaped failure rate. Hence,
the WBS distribution is appropriate for modeling this data set.\medskip

\noindent Table 1 lists the MLEs and their corresponding standard errors in
parentheses of parameters of the WBS, BBS, KBS, McBS, MOEBS, GBS, EBS and BS
distributions for Meeker and Escobar data set. The statistics $-2\widehat{\ell
}$, AIC, BIC, CAIC, K-S and its p-value are listed in Table 2 for all the
distributions. These results show that the WBS distribution has the largest
p-value and the smallest $-2\widehat{\ell}$, AIC, BIC, CAIC and K-S values.
So, the WBS distribution gives an excellent fit than the others models for
Meeker and Escobar data set. The histogram of this data set and the plots of
the estimated densities of all models are given in Figure 6. From this Figure,
we can conclude that the WBS model provides a better fit to the histogram and
therefore could be chosen as the best model for Meeker and Escobar
data.\bigskip

\subsection{Turbochargers failure data}

The second data set represents the time to failure($10^{3}$ h) of turbocharger
of one type of engine given in Xu et al. \cite{r27}. The data set is: 1.6,
2.0, 2.6, 3.0, 3.5, 3.9, 4.5, 4.6, 4.8, 5.0, 5.1, 5.3, 5.4, 5.6, 5.8, 6.0,
6.0, 6.1, 6.3, 6.5, 6.5, 6.7, 7.0, 7.1, 7.3, 7.3, 7.3, 7.7, 7.7, 7.8, 7.9,
8.0, 8.1, 8.3, 8.4, 8.4, 8.5, 8.7, 8.8, 9.0. The Figure 5(b) shows concave TTT
plot for the data set, indicating increasing failure rate function, which can
be properly accommodated by a WBS distribution.\medskip

\noindent Table 3 gives the MLEs of the parameters of all models used here and
their corresponding standard errors in parentheses. The statistics
$-2\widehat{\ell}$, AIC, BIC, CAIC, K-S and its p-value are listed in Table 4.
From this Table , we can see the WBS distribution as the best fit for the data
set among all the seven models. The histogram of this data set and the plots
of the estimated densities of all models are shown in Figure 7. So, the WBS
model provides a better fit to second data set.\bigskip%

%TCIMACRO{\TeXButton{B}{\begin{table}[h] \centering}}%
%BeginExpansion
\begin{table}[h] \centering
%EndExpansion
\caption{MLEs and their standard errors in parentheses for the first data.}%
\begin{tabular}
[c]{lccccc}\hline
Model & $\alpha$ & $\beta$ & $a$ & $b$ & $c$\\\hline
WBS & $0.8152$ & $22.9053$ & $0.1115$ & $0.2683$ & $-$\\
& $\left(  0.5466\right)  $ & $\left(  13.5555\right)  $ & $\left(
0.0674\right)  $ & $\left(  0.2193\right)  $ & $-$\\
McBS & $22.3663$ & $0.3293$ & $6.9147$ & $124.9055$ & $67.0133$\\
& $\left(  19.2479\right)  $ & $\left(  0.5770\right)  $ & $\left(
1.3436\right)  $ & $\left(  153.8349\right)  $ & $\left(  41.2595\right)  $\\
MOEBS & $1.9735$ & $13.8678$ & $17.1905$ & $-$ & $-$\\
& $\left(  0.5232\right)  $ & $\left(  7.7663\right)  $ & $\left(
10.0909\right)  $ & $-$ & $-$\\
KBS & $11.3624$ & $6.5795$ & $11.1898$ & $72.6776$ & $-$\\
& $\left(  6.0551\right)  $ & $\left(  7.5249\right)  $ & $\left(
2.3506\right)  $ & $\left(  102.1419\right)  $ & $-$\\
GBS & $5.6073$ & $1.3777$ & $3.6317$ & $-$ & $-$\\
& $\left(  1.8869\right)  $ & $\left(  0.9472\right)  $ & $\left(
0.5234\right)  $ & $-$ & $-$\\
BBS & $15.6640$ & $3.9207$ & $31.7249$ & $17.4625$ & $-$\\
& $\left(  17.9647\right)  $ & $\left(  5.5777\right)  $ & $\left(
49.6707\right)  $ & $\left(  36.5502\right)  $ & $-$\\
EBS & $4.8477$ & $3.8141$ & $5.7211$ & $-$ & $-$\\
& $\left(  3.0942\right)  $ & $\left(  5.1989\right)  $ & $\left(
1.8078\right)  $ & $-$ & $-$\\
BS & $1.6778$ & $64.0791$ & $-$ & $-$ & $-$\\
& $\left(  0.2218\right)  $ & $\left(  14.5028\right)  $ & $-$ & $-$ &
$-$\\\hline
\end{tabular}%
%TCIMACRO{\TeXButton{E}{\end{table}}}%
%BeginExpansion
\end{table}%
%EndExpansion
%

%TCIMACRO{\TeXButton{B}{\begin{table}[h] \centering}}%
%BeginExpansion
\begin{table}[h] \centering
%EndExpansion
\caption{MLEs and their standard errors in parentheses for the second data.}%
\begin{tabular}
[c]{lccccc}\hline
Model & $\alpha$ & $\beta$ & $a$ & $b$ & $c$\\\hline
WBS & $0.2007$ & $3.4802$ & $0.1185$ & $0.2323$ & $-$\\
& $\left(  0.127081\right)  $ & $\left(  0.626594\right)  $ & $\left(
0.072865\right)  $ & $\left(  0.255233\right)  $ & $-$\\
McBS & $10.8469$ & $0.0311$ & $21.5229$ & $51.3146$ & $59.8247$\\
& $\left(  0.3520032\right)  $ & $\left(  0.0016003\right)  $ & $\left(
4.1772473\right)  $ & $\left(  0.1028498\right)  $ & $\left(
0.1186039\right)  $\\
MOEBS & $0.5269$ & $2.1087$ & $74.3785$ & $-$ & $-$\\
& $\left(  0.1009906\right)  $ & $\left(  0.4566437\right)  $ & $\left(
0.0049525\right)  $ & $-$ & $-$\\
KBS & $7.7703$ & $0.1109$ & $23.9059$ & $63.4929$ & $-$\\
& $\left(  2.8922\right)  $ & $\left(  0.0834\right)  $ & $\left(
3.4606\right)  $ & $\left(  0.0374\right)  $ & $-$\\
GBS & $4.5864$ & $0.0160$ & $11.6225$ & $-$ & $-$\\
& $\left(  0.0924\right)  $ & $\left(  0.0027\right)  $ & $\left(
1.9661\right)  $ & $-$ & $-$\\
BBS & $10.9655$ & $0.0655$ & $64.5533$ & $15.7442$ & $-$\\
& $\left(  11.8708\right)  $ & $\left(  0.1504\right)  $ & $\left(
1.9297\right)  $ & $\left(  4.1212\right)  $ & $-$\\
EBS & $5.1071$ & $0.0493$ & $41.1747$ & $-$ & $-$\\
& $\left(  12.3684\right)  $ & $\left(  0.2397\right)  $ & $\left(
17.0774\right)  $ & $-$ & $-$\\
BS & $0.4139$ & $5.7538$ & $-$ & $-$ & $-$\\
& $\left(  0.0463\right)  $ & $\left(  0.3684\right)  $ & $-$ & $-$ &
$-$\\\hline
\end{tabular}%
%TCIMACRO{\TeXButton{E}{\end{table}}}%
%BeginExpansion
\end{table}%
%EndExpansion
%

%TCIMACRO{\TeXButton{B}{\begin{table}[h] \centering}}%
%BeginExpansion
\begin{table}[h] \centering
%EndExpansion
\caption{The statistics: -2$\hat \ell $, AIC, CAIC, K-S and its p-value for
the first data.}%
\begin{tabular}
[c]{lcccccc}\hline
Model & $-2\widehat{\ell}$ & AIC & BIC & CAIC & K-S & p-Value\\\hline
WBS & $352.8431$ & $360.8431$ & $366.4479$ & $362.4431$ & $0.1685$ &
$0.3618$\\
McBS & $357.8659$ & $367.8659$ & $374.8719$ & $370.3659$ & $0.2273$ & $0.0899
$\\
MOEBS & $363.1652$ & $369.1652$ & $373.3688$ & $370.0883$ & $0.1854$ &
$0.2536$\\
KBS & $362.5005$ & $370.5005$ & $376.1053$ & $372.1005$ & $0.2054$ &
$0.1591$\\
BBS & $366.5623$ & $374.5623$ & $380.1671$ & $376.1623$ & $0.2075$ &
$0.1510$\\
GBS & $367.4188$ & $373.4188$ & $377.6223$ & $374.3418$ & $0.2213$ &
$0.1058$\\
EBS & $368.9539$ & $374.9539$ & $379.1575$ & $375.8770$ & $0.2043$ &
$0.1635$\\
BS & $385.5103$ & $389.5103$ & $392.3127$ & $389.9547$ & $0.3218$ &
$0.0040$\\\hline
\end{tabular}%
%TCIMACRO{\TeXButton{E}{\end{table}}}%
%BeginExpansion
\end{table}%
%EndExpansion

\begin{center}%
%TCIMACRO{\TeXButton{B}{\begin{table}[h] \centering}}%
%BeginExpansion
\begin{table}[h] \centering
%EndExpansion
\caption{The statistics: -2$\hat \ell $, AIC, CAIC, K-S and its p-value for
the second data.}%
\begin{tabular}
[c]{lcccccc}\hline
Model & $-2\widehat{\ell}$ & AIC & BIC & CAIC & K-S & p-Value\\\hline
WBS & $157.1875$ & $165.1875$ & $171.9431$ & $166.3304$ & $0.0778$ &
$0.9685$\\
McBS & $164.9313$ & $174.9313$ & $183.3757$ & $176.696$ & $0.1066$ &
$0.7535$\\
MOEBS & $167.0805$ & $173.0805$ & $178.1472$ & $173.7472$ & $0.0909$ &
$0.8958$\\
KBS & $166.1958$ & $174.1958$ & $180.9513$ & $175.3387$ & $0.1119$ &
$0.6976$\\
BBS & $173.0616$ & $181.0616$ & $187.8172$ & $182.2045$ & $0.1205$ &
$0.6067$\\
GBS & $173.4768$ & $179.4768$ & $184.5434$ & $180.1435$ & $0.1199$ &
$0.6136$\\
EBS & $180.8146$ & $186.8146$ & $191.8812$ & $187.4813$ & $0.1607$ &
$0.2531$\\
BS & $182.7348$ & $186.7348$ & $190.1125$ & $187.0591$ & $0.1653$ &
$0.2243$\\\hline
\end{tabular}%
%TCIMACRO{\TeXButton{E}{\end{table}}}%
%BeginExpansion
\end{table}%
%EndExpansion

\end{center}

%

%TCIMACRO{\FRAME{itbpFU}{5.4743in}{2.7752in}{0in}{\Qcb{Figure 5. TTT-transform
%plot for the (a) the first data; (b) second data.}}{}{fig5.eps}%
%{\special{ language "Scientific Word";  type "GRAPHIC";  display "USEDEF";
%valid_file "F";  width 5.4743in;  height 2.7752in;  depth 0in;
%original-width 18.9204in;  original-height 9.1584in;  cropleft "0";
%filename 'fig5.eps';file-properties "XNPEU";}} }%
%BeginExpansion
{\parbox[b]{5.4743in}{\begin{center}
\ifcase\msipdfoutput
\includegraphics[
height=2.7752in,
width=5.4743in
]%
{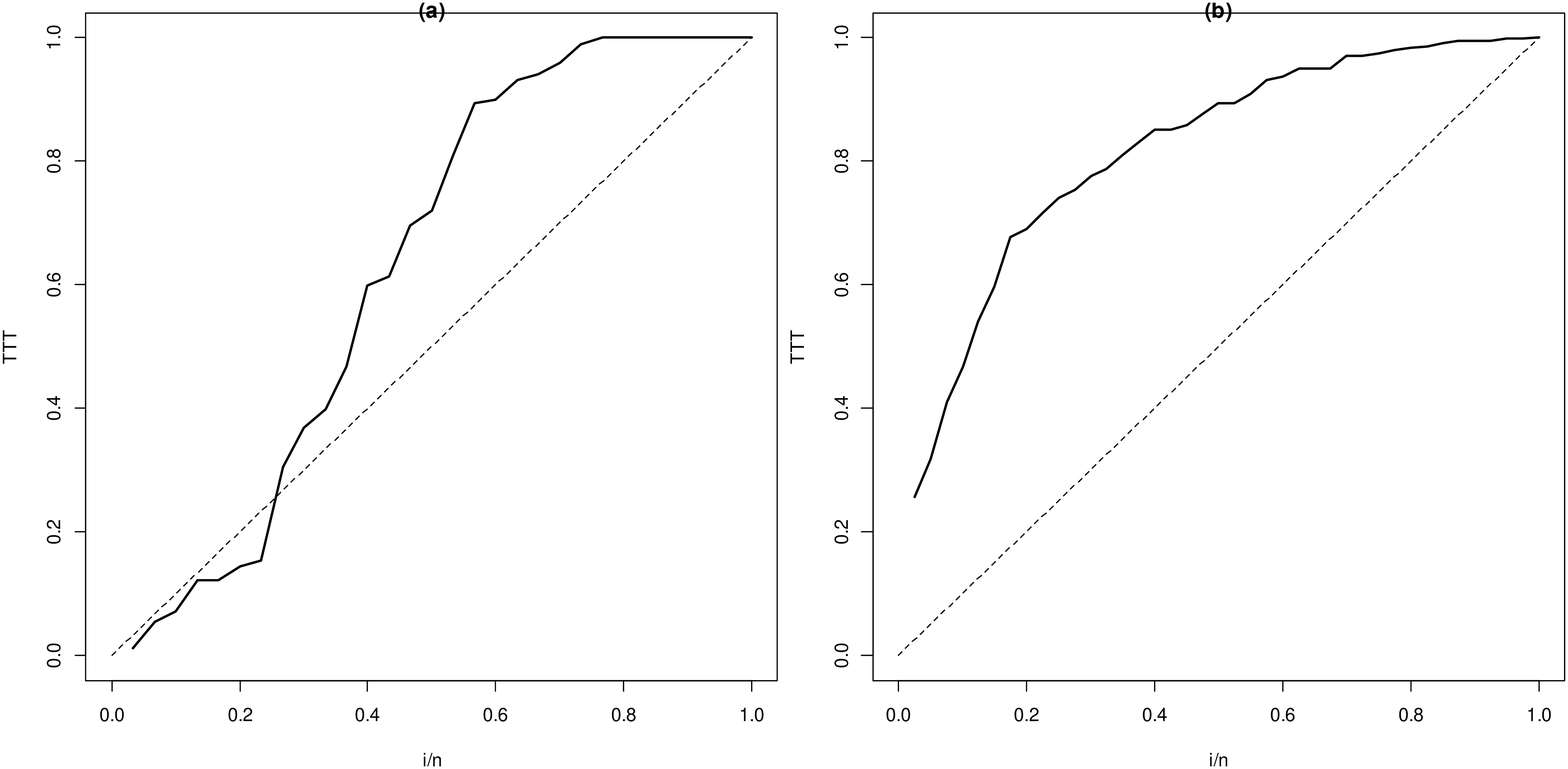}%
\else
\includegraphics[
height=2.7752in,
width=5.4743in
]%
{C:/Users/LAZHAR/IDEES/Weibull-Birnbaum-Saunders distribution/arXiv/graphics/fig5.jpg}%
\fi
\\
Figure 5. TTT-transform plot for the (a) the first data; (b) second data.
\end{center}}}
%EndExpansion
%

%TCIMACRO{\FRAME{itbpFU}{5.2857in}{2.6109in}{0in}{\Qcb{Figure 6. The histogram
%and the estimated densities of the first data set.}}{}{fig6.eps}%
%{\special{ language "Scientific Word";  type "GRAPHIC";  display "USEDEF";
%valid_file "F";  width 5.2857in;  height 2.6109in;  depth 0in;
%original-width 18.9204in;  original-height 9.1584in;  cropleft "0";
%croptop "0.9083";  cropright "1";  cropbottom "0";
%filename 'fig6.eps';file-properties "XNPEU";}} }%
%BeginExpansion
{\parbox[b]{5.2857in}{\begin{center}
\ifcase\msipdfoutput
\includegraphics[
trim=0.000000in 0.000000in 0.000000in 0.839825in,
height=2.6109in,
width=5.2857in
]%
{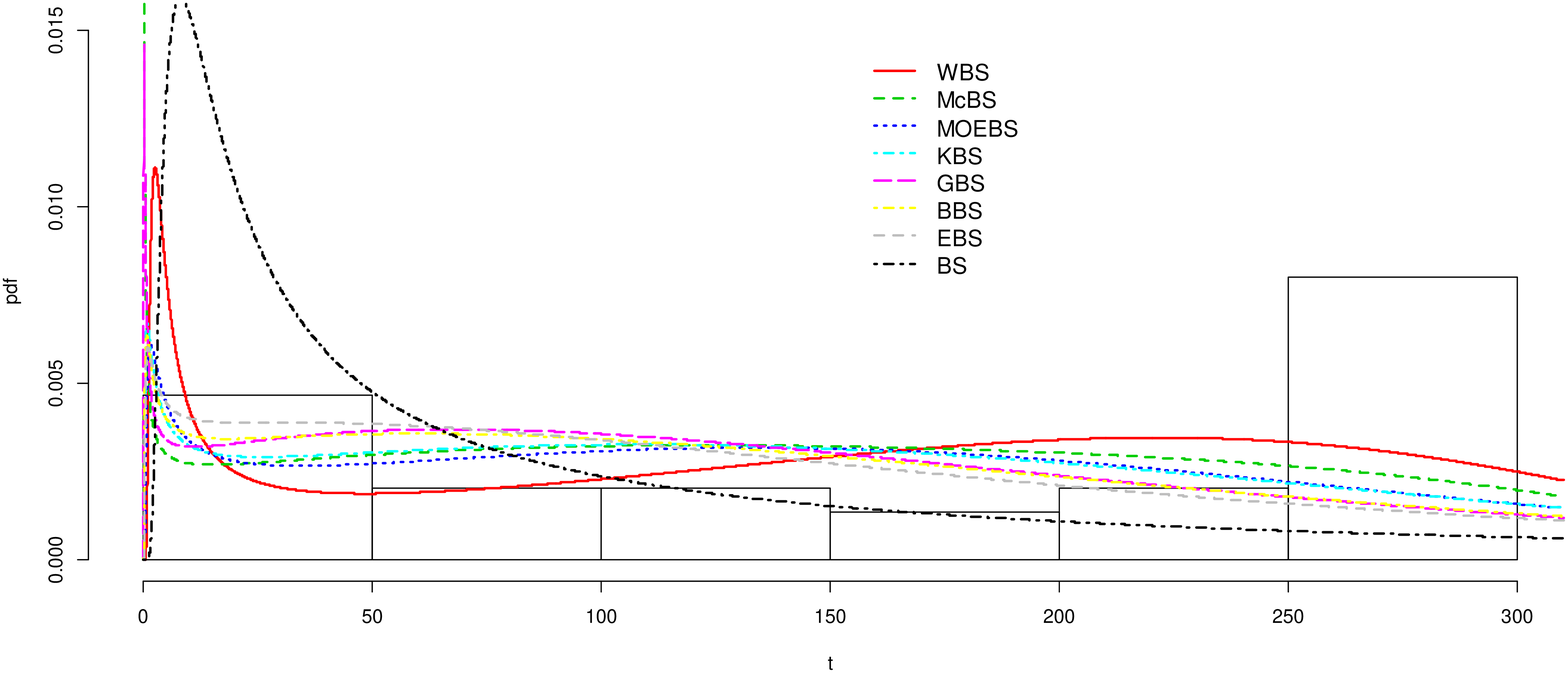}%
\else
\includegraphics[
height=2.6109in,
width=5.2857in
]%
{C:/Users/LAZHAR/IDEES/Weibull-Birnbaum-Saunders distribution/arXiv/graphics/fig6.jpg}%
\fi
\\
Figure 6. The histogram and the estimated densities of the first data set.
\end{center}}}
%EndExpansion
%

%TCIMACRO{\FRAME{itbpFU}{5.2442in}{2.6195in}{0in}{\Qcb{Figure 7. The histogram
%and the estimated densities of the second data set.}}{}{fig7.eps}%
%{\special{ language "Scientific Word";  type "GRAPHIC";  display "USEDEF";
%valid_file "F";  width 5.2442in;  height 2.6195in;  depth 0in;
%original-width 13.4089in;  original-height 8.6438in;  cropleft "0";
%croptop "0.8818";  cropright "0.9718";  cropbottom "0";
%filename 'fig7.eps';file-properties "XNPEU";}} }%
%BeginExpansion
{\parbox[b]{5.2442in}{\begin{center}
\ifcase\msipdfoutput
\includegraphics[
trim=0.000000in 0.000000in 0.378131in 1.021697in,
height=2.6195in,
width=5.2442in
]%
{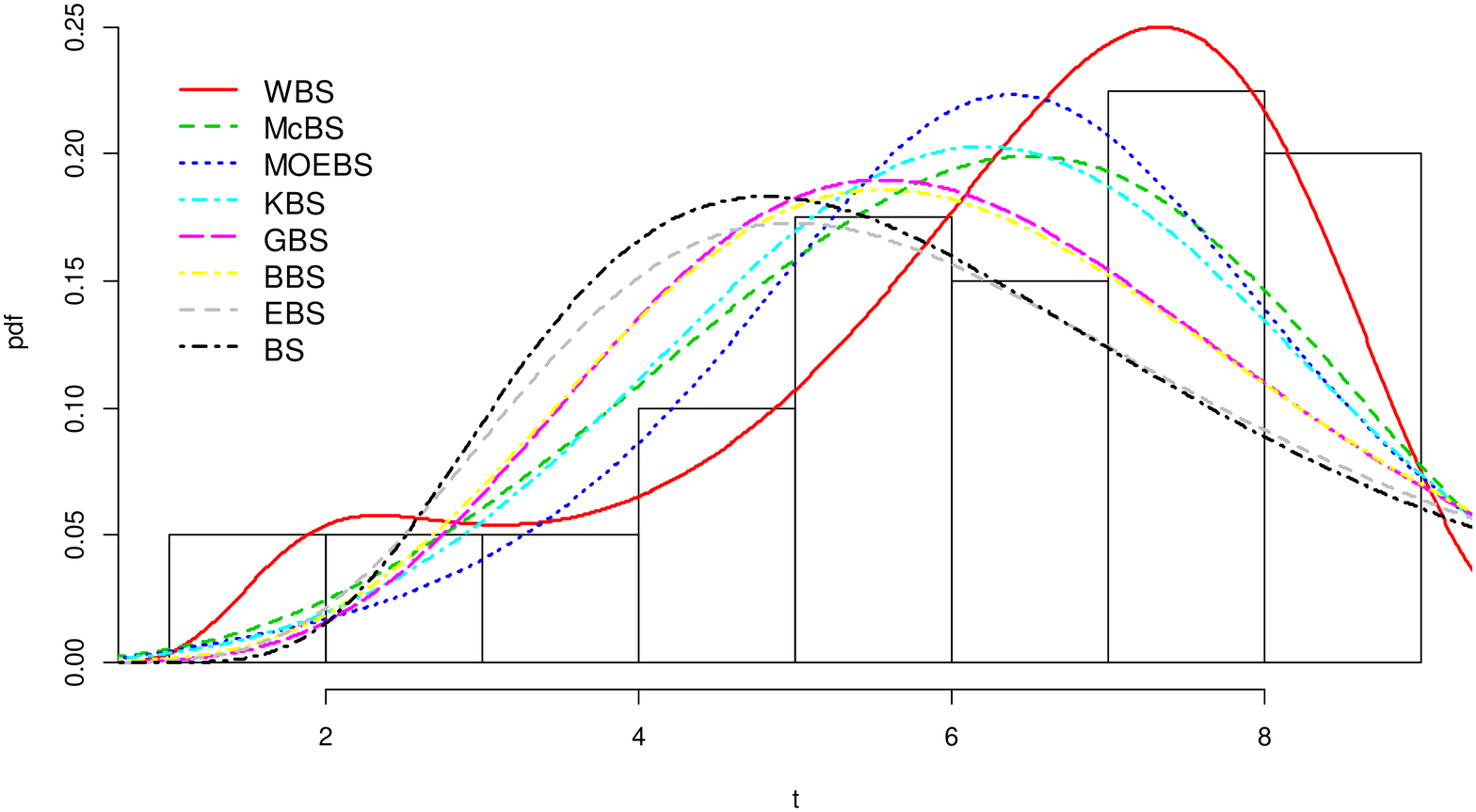}%
\else
\includegraphics[
height=2.6195in,
width=5.2442in
]%
{C:/Users/LAZHAR/IDEES/Weibull-Birnbaum-Saunders distribution/arXiv/graphics/fig7.jpg}%
\fi
\\
Figure 7. The histogram and the estimated densities of the second data set.
\end{center}}}
%EndExpansion

\bigskip

\section{\textbf{Conclusions\label{sec7}}}

In this paper, we introduce a new four-parameter model called the Weibull
Birnbaum-Saunders (WBS) distribution that extends the Birnbaum-Saunders
distribution. The WBS hazard function can be
decreasing,\ increasing,\textbf{\ }upside-down bathtub, bathtub-shaped or
modified bathtub shaped depending on its parameters. So, the WBS model can be
used quite effectively in analyzing lifetime data. The properties of the new
distribution including expansions for the density function, moments,
generating function, order statistics, quantile function, mean deviations and
reliability are provided. We estimate the model parameters by maximum
likelihood and obtain the observed information matrix. An application of the
WBS distribution to two real data sets is used to illustrate that this
distribution provides a better fit than many other related non-nested
models.\bigskip

\section{\textbf{Appendix}}

The elements of the observed information matrix $J\left(  \mathbf{\xi}\right)
$ for the parameters $\left(  \alpha,\beta,a,b\right)  $ are%
\begin{align*}
L_{\alpha\alpha}  &  =\frac{n}{\alpha^{2}}+\frac{6n}{\alpha^{4}}-\frac
{3}{\alpha^{4}}\sum_{i=1}^{n}\left(  \frac{t_{i}}{\beta}+\frac{\beta}{t_{i}%
}\right)  +\frac{2\left(  b-1\right)  }{\alpha^{2}}\sum_{i=1}^{n}\frac
{v_{i}\phi\left(  v_{i}\right)  }{\Phi\left(  v_{i}\right)  }\\
&  -\frac{2\left(  b+1\right)  }{\alpha^{2}}\sum_{i=1}^{n}\frac{v_{i}%
\phi\left(  v_{i}\right)  }{1-\Phi\left(  v_{i}\right)  }+\frac{\left(
b-1\right)  }{\alpha^{3}}\sum_{i=1}^{n}\left\{  \frac{v_{i}^{4}\phi\left(
v_{i}\right)  }{\Phi\left(  v_{i}\right)  }-\frac{\alpha v_{i}^{2}\phi
^{2}\left(  v_{i}\right)  }{\Phi^{2}\left(  v_{i}\right)  }\right\} \\
&  -\frac{\left(  b+1\right)  }{\alpha^{3}}\sum_{i=1}^{n}\left\{  \frac
{v_{i}^{4}\phi\left(  v_{i}\right)  }{1-\Phi\left(  v_{i}\right)  }%
-\frac{\alpha v_{i}^{2}\phi^{2}\left(  v_{i}\right)  }{\left[  1-\Phi\left(
v_{i}\right)  \right]  ^{2}}\right\}  +\frac{ab}{\alpha^{2}}\sum_{i=1}%
^{n}\frac{v_{i}\phi\left(  v_{i}\right)  \Phi^{b-1}\left(  v_{i}\right)
}{\left[  1-\Phi\left(  v_{i}\right)  \right]  ^{b+1}}\\
&  +\frac{ab}{\alpha^{2}}\sum_{i=1}^{n}\frac{v_{i}\left(  v_{i}^{2}-1\right)
\phi\left(  v_{i}\right)  \Phi^{b-1}\left(  v_{i}\right)  }{\left[
1-\Phi\left(  v_{i}\right)  \right]  ^{b+1}}+\left(  b-1\right)  \sum
_{i=1}^{n}\frac{v_{i}\phi\left(  v_{i}\right)  ^{2}\Phi^{b-2}\left(
v_{i}\right)  }{\left[  1-\Phi\left(  v_{i}\right)  \right]  ^{b+1}}\\
&  +\left(  b+1\right)  \sum_{i=1}^{n}\frac{v_{i}\phi^{2}\left(  v_{i}\right)
\Phi^{b-1}\left(  v_{i}\right)  }{\left[  1-\Phi\left(  v_{i}\right)  \right]
^{b+2}},\\
L_{\alpha\beta}  &  =-\frac{1}{\alpha^{3}\beta}\sum_{i=1}^{n}\left(
\frac{t_{i}}{\beta}-\frac{\beta}{t_{i}}\right)  +\frac{\left(  b-1\right)
}{2\beta\alpha^{2}}\sum_{i=1}^{n}\left\{  \frac{\alpha v_{i}\phi\left(
v_{i}\right)  }{\Phi\left(  v_{i}\right)  }+\frac{v_{i}^{4}\phi\left(
v_{i}\right)  }{\Phi\left(  v_{i}\right)  }-\frac{\alpha v_{i}^{2}\phi
^{2}\left(  v_{i}\right)  }{\Phi^{2}\left(  v_{i}\right)  }\right\} \\
&  -\frac{\left(  b-1\right)  }{2\beta\alpha^{2}}\sum_{i=1}^{n}\left\{
\frac{\alpha v_{i}\phi\left(  v_{i}\right)  }{1-\Phi\left(  v_{i}\right)
}+\frac{v_{i}^{4}\phi\left(  v_{i}\right)  }{1-\Phi\left(  v_{i}\right)
}-\frac{\alpha v_{i}^{2}\phi^{2}\left(  v_{i}\right)  }{\left[  1-\Phi\left(
v_{i}\right)  \right]  ^{2}}\right\} \\
&  +\frac{ab}{2\beta\alpha^{2}}\sum_{i=1}^{n}\frac{\tau\left(  \sqrt
{t_{i}/\beta}\right)  \phi\left(  v_{i}\right)  \left(  v_{i}^{2}\phi\left(
v_{i}\right)  -1\right)  \Phi^{b-1}\left(  v_{i}\right)  }{\left[
1-\Phi\left(  v_{i}\right)  \right]  ^{b+1}}\\
&  -\frac{ab\left(  b-1\right)  }{2\beta\alpha^{2}}\sum_{i=1}^{n}\frac
{\tau\left(  \sqrt{t_{i}/\beta}\right)  \phi\left(  v_{i}\right)  \Phi
^{b-2}\left(  v_{i}\right)  }{\left[  1-\Phi\left(  v_{i}\right)  \right]
^{b+1}}\\
&  -\frac{ab\left(  b+1\right)  }{2\beta\alpha^{2}}\sum_{i=1}^{n}\frac
{\tau\left(  \sqrt{t_{i}/\beta}\right)  v_{i}\phi^{2}\left(  v_{i}\right)
\Phi^{b-1}\left(  v_{i}\right)  }{\left[  1-\Phi\left(  v_{i}\right)  \right]
^{b+2}},
\end{align*}

\begin{align*}
L_{\beta\beta}  &  =\frac{n}{2\beta^{2}}-\sum_{i=1}^{n}\frac{1}{\left(
t_{i}+\beta\right)  ^{2}}-\frac{1}{\alpha^{2}\beta^{3}}\sum_{i=1}^{n}%
t_{i}+\frac{\left(  b-1\right)  }{2\alpha\beta^{2}}\sum_{i=1}^{n}\frac
{\tau\left(  \sqrt{t_{i}/\beta}\right)  \phi\left(  v_{i}\right)  }%
{\Phi\left(  v_{i}\right)  }\\
&  -\frac{\left(  b-1\right)  }{4\alpha\beta^{2}}\sum_{i=1}^{n}\left\{
-\frac{\alpha v_{i}\phi\left(  v_{i}\right)  }{\Phi\left(  v_{i}\right)
}+\frac{v_{i}\tau^{2}\left(  \sqrt{t_{i}/\beta}\right)  \phi\left(
v_{i}\right)  }{\alpha\Phi\left(  v_{i}\right)  }+\frac{v_{i}\tau^{2}\left(
\sqrt{t_{i}/\beta}\right)  \phi^{2}\left(  v_{i}\right)  }{\alpha\Phi
^{2}\left(  v_{i}\right)  }\right\} \\
&  +\frac{\left(  b+1\right)  }{4\alpha\beta^{2}}\sum_{i=1}^{n}\left\{
-\frac{\alpha v_{i}\phi\left(  v_{i}\right)  }{1-\Phi\left(  v_{i}\right)
}+\frac{v_{i}\tau^{2}\left(  \sqrt{t_{i}/\beta}\right)  \phi\left(
v_{i}\right)  }{\alpha\left[  1-\Phi\left(  v_{i}\right)  \right]  }%
-\frac{v_{i}\tau^{2}\left(  \sqrt{t_{i}/\beta}\right)  \phi^{2}\left(
v_{i}\right)  }{\alpha\left[  1-\Phi\left(  v_{i}\right)  \right]  ^{2}%
}\right\} \\
&  -\frac{ab}{2\alpha\beta^{2}}\sum_{i=1}^{n}\frac{\tau\left(  \sqrt
{t_{i}/\beta}\right)  \phi\left(  v_{i}\right)  \Phi\left(  v_{i}\right)
^{b-1}}{\left[  1-\Phi\left(  v_{i}\right)  \right]  ^{b+1}}-\frac{\left(
b+1\right)  }{2\alpha\beta^{2}}\sum_{i=1}^{n}\frac{\tau\left(  \sqrt
{t_{i}/\beta}\right)  \phi\left(  v_{i}\right)  }{1-\Phi\left(  v_{i}\right)
}\\
&  +\frac{ab}{2\alpha^{2}\beta^{2}}\sum_{i=1}^{n}\frac{\phi\left(
v_{i}\right)  \Phi^{b-1}\left(  v_{i}\right)  }{\left[  1-\Phi\left(
v_{i}\right)  \right]  ^{b+1}}\left\{  \tau^{2}\left(  \sqrt{t_{i}/\beta
}\right)  v_{i}\phi\left(  v_{i}\right)  -\sqrt{t_{i}/\beta}+\sqrt{\beta
/t_{i}}\right\} \\
&  -\frac{ab\left(  b-1\right)  }{2\alpha^{2}\beta^{2}}\sum_{i=1}^{n}%
\frac{\tau^{2}\left(  \sqrt{t_{i}/\beta}\right)  \phi^{2}\left(  v_{i}\right)
\Phi^{b-1}\left(  v_{i}\right)  }{\left[  1-\Phi\left(  v_{i}\right)  \right]
^{b+1}}\\
&  +\frac{ab\left(  b+1\right)  }{2\alpha^{2}\beta^{2}}\sum_{i=1}^{n}%
\frac{\tau^{2}\left(  \sqrt{t_{i}/\beta}\right)  \phi^{2}\left(  v_{i}\right)
\Phi^{b-1}\left(  v_{i}\right)  }{\left[  1-\Phi\left(  v_{i}\right)  \right]
^{b+2}},
\end{align*}

\bigskip%
\begin{align*}
L_{\beta b}  &  =\frac{1}{2\beta\alpha}\sum_{i=1}^{n}\frac{\tau\left(
\sqrt{t_{i}/\beta}\right)  \phi\left(  v_{i}\right)  }{\Phi\left(
v_{i}\right)  }+\frac{1}{2\beta\alpha}\sum_{i=1}^{n}\frac{\tau\left(
\sqrt{t_{i}/\beta}\right)  \phi\left(  v_{i}\right)  }{1-\Phi\left(
v_{i}\right)  }\\
&  -\frac{1}{2\beta\alpha}\sum_{i=1}^{n}\frac{\tau\left(  \sqrt{t_{i}/\beta
}\right)  \phi\left(  v_{i}\right)  \Phi^{b-1}\left(  v_{i}\right)  }{\left[
1-\Phi\left(  v_{i}\right)  \right]  ^{b+1}}\left\{  1+b\log\left[  \frac
{\Phi\left(  v_{i}\right)  }{1-\Phi\left(  v_{i}\right)  }\right]  \right\}
,\\
L_{\alpha b}  &  =\frac{1}{\alpha}\sum_{i=1}^{n}\frac{v_{i}\phi\left(
v_{i}\right)  }{\Phi\left(  v_{i}\right)  }+\frac{1}{\alpha}\sum_{i=1}%
^{n}\frac{v_{i}\phi\left(  v_{i}\right)  }{1-\Phi\left(  v_{i}\right)  }\\
&  -\frac{a}{\alpha}\sum_{i=1}^{n}\frac{v_{i}\phi\left(  v_{i}\right)
\Phi^{b-1}\left(  v_{i}\right)  }{\left[  1-\Phi\left(  v_{i}\right)  \right]
^{b+1}}\left\{  1+b\log\left[  \frac{\Phi\left(  v_{i}\right)  }{1-\Phi\left(
v_{i}\right)  }\right]  \right\}  ,
\end{align*}

\begin{align*}
L_{\beta a}  &  =\frac{b}{2\beta\alpha}\sum_{i=1}^{n}\frac{\tau\left(
\sqrt{t_{i}/\beta}\right)  \phi\left(  v_{i}\right)  \Phi^{b-1}\left(
v_{i}\right)  }{\left[  1-\Phi\left(  v_{i}\right)  \right]  ^{b+1}},\text{
}L_{\alpha a}=\frac{b}{\alpha}\sum_{i=1}^{n}\frac{v_{i}\phi\left(
v_{i}\right)  \Phi^{b-1}\left(  v_{i}\right)  }{\left[  1-\Phi\left(
v_{i}\right)  \right]  ^{b+1}},\\
L_{bb}  &  =-\frac{n}{b^{2}}-a\sum_{i=1}^{n}\left[  \frac{\Phi\left(
v_{i}\right)  }{1-\Phi\left(  v_{i}\right)  }\right]  ^{b}\left(  \log\left[
\frac{\Phi\left(  v_{i}\right)  }{1-\Phi\left(  v_{i}\right)  }\right]
\right)  ^{2},\\
L_{aa}  &  =-\frac{n}{a^{2}}\text{ \ and \ }L_{ab}=-\sum_{i=1}^{n}\left[
\frac{\Phi\left(  v_{i}\right)  }{1-\Phi\left(  v_{i}\right)  }\right]
^{b}\log\left[  \frac{\Phi\left(  v_{i}\right)  }{1-\Phi\left(  v_{i}\right)
}\right]  .
\end{align*}

\bigskip

\bigskip

\bigskip

\bigskip
\end{document}